\documentclass[lettersize,journal]{IEEEtran}

\usepackage[T1]{fontenc}
\usepackage[utf8]{inputenc}

\usepackage{amsmath,amssymb,amsfonts}
\usepackage{algorithm}
\usepackage{algpseudocode}
\usepackage{graphicx}
\usepackage{booktabs}
\usepackage{array}
\usepackage{textcomp}
\usepackage{stfloats}
\usepackage{url}
\usepackage{cite}
\usepackage[hidelinks,bookmarks=false]{hyperref}

\makeatletter
\def\@IEEENORMtitlevspace{1.5\baselineskip}
\def\@IEEEMINtitlevspace{1\baselineskip}
\def\IEEEtitletopspaceextra{5pt}
\def\tablename{Table}
\long\def\@makecaption#1#2{%
\ifx\@captype\@IEEEtablestring
\par
\ifdim\prevdepth=-1000pt\def\@tabcapstrut{\rule{0pt}{\topskip}}\else\let\@tabcapstrut\relax\fi
\setbox\@tempboxa\hbox{\normalfont\small\@tabcapstrut{#1.}\nobreakspace\nobreakspace #2}%
\ifdim \wd\@tempboxa >\hsize
\setbox\@tempboxa\hbox{\normalfont\small\@tabcapstrut{#1.}\nobreakspace\nobreakspace}%
\parbox[t]{\hsize}{\normalfont\small\noindent\unhbox\@tempboxa#2}%
\else
\hbox to\hsize{\normalfont\small\hfil\box\@tempboxa\hfil}%
\fi
{\footnotesize\par\addvspace{0.5\baselineskip}}%
\@IEEEtablecaptionsepspace
\else
\@IEEEfigurecaptionsepspace
\setbox\@tempboxa\hbox{\normalfont\small {#1.}\nobreakspace\nobreakspace #2}%
\ifdim \wd\@tempboxa >\hsize
\setbox\@tempboxa\hbox{\normalfont\small {#1.}\nobreakspace\nobreakspace}%
\parbox[t]{\hsize}{\normalfont\small\noindent\unhbox\@tempboxa#2}%
\else
\hbox to\hsize{\normalfont\small\box\@tempboxa\hfil}%
\fi\fi}
\makeatother

\begin{document}

\title{Look It Up: Multi-bit LLM Watermarking\\ via Codebook Search}

\author{Jaros{\l}aw~Janas\textsuperscript{1},
        Pawe{\l}~Morawiecki\textsuperscript{1},
        and~Josef~Pieprzyk\textsuperscript{1,2}\\[1.2ex]
{\small\itshape\textsuperscript{\upshape 1}Institute of Computer Science, Polish Academy of Sciences, 01-248 Warsaw, Poland\\
\textsuperscript{\upshape 2}CSIRO, Marsfield, NSW 2122, Australia\\[\dimexpr0.5ex+6.8pt\relax]
\upshape Corresponding e-mail: \texttt{jaroslawjanas.dev@gmail.com}\\[-3.4pt]
Code: \href{https://github.com/jaroslawjanas/llmw-codebook}{\texttt{https://github.com/jaroslawjanas/llmw-codebook}}}\\[2.4ex]
\parbox{0.8\textwidth}{\centering\footnotesize
This work was supported in part by the National Science Centre, Poland, under Grant OPUS 2023/49/B/ST6/02580.\\[0.5ex]
This work has been submitted to the IEEE for possible publication. Copyright may be transferred without notice, after which this version may no longer be accessible.}}

\markboth{}%
{Janas \MakeLowercase{\textit{et al.}}: Look It Up: Multi-bit LLM Watermarking via Codebook Search}

\maketitle

\begin{abstract}
Attributing short LLM-generated texts to a specific source remains challenging for existing multi-bit watermarking methods, which often require long generations or costly decoding to recover payloads reliably. We propose a codebook-based watermarking scheme that maps an $n$-bit payload to a longer $u$-bit codeword and embeds it during generation through token-level logit biasing. During extraction, the received bit sequence is matched to the nearest codeword in a precomputed secret codebook using Hamming or Levenshtein distance. Experiments show that our method achieves near-perfect recovery for 16--24 bit payloads on 200-token generations and strong performance on 127-token generations for base models. Results are weaker on some instruction-tuned models with very short texts, where low-entropy output distributions reduce embedding reliability. We also study a BCH-based variant that avoids codebook storage but generally reduces robustness. The Levenshtein variant improves resilience to deletions and light paraphrasing, while insertions and stronger rewrites remain challenging. We analyze collision probability, codeword separation, extraction time, tokenization consistency, and text quality, highlighting both the practicality and limitations of codebook-based watermarking for short-form attribution.
\end{abstract}


\section{Introduction}
\label{sec:introduction}
\IEEEPARstart{L}{arge} Language Models (LLMs) have become widely adopted for tasks ranging from content drafting and summarisation to dialogue and code generation. As the quality of machine-generated text has improved, so has the difficulty of distinguishing it from human writing---raising concerns about misuse in misinformation, plagiarism, and large-scale deceptive content such as fake product reviews. Early watermarking methods addressed the detection problem: zero-bit schemes \cite{kirchenbauer2023,wang2024} embed a statistical signal that allows a verifier to determine whether a given text was produced by an LLM, but they do not carry any identifying information about the source.

Multi-bit watermarking goes further by embedding a payload---such as a user or session identifier---directly into the generated token sequence, enabling attribution to a specific source. Several such schemes have since been proposed (see Section~\ref{sec:current_approaches}), yet they are predominantly evaluated on texts of approximately 200 tokens, as longer sequences provide more redundancy for reliable extraction. This leaves the sub-200 token space largely unaddressed. Product reviews, social media posts, and short comments---all common targets for AI-generated deception---routinely fall well below this threshold. Studies estimate that 16--33\% of online reviews are inauthentic \cite{sahut2024}, with recent analyses putting this figure at roughly 30\% as of 2025, a problem projected to cost hundreds of billions of dollars globally \cite{capitalone2025}. Reliably watermarking these short-form texts remains an open and practically important challenge.

\section{Current Approaches and Limitations}
\label{sec:current_approaches}

Several recent works have explored multi-bit watermarking schemes for LLM-generated text beyond the simple zero-bit (detector) watermarks \cite{wang2024,yoo2024,qu2025,fernandez2023,lin2024dermark}. Table~\ref{tab:baselines} summarizes the performance of representative methods from the literature. Overall, there is a clear trade-off between accuracy and efficiency in existing approaches.

\begin{table*}[ht]
\caption{Comparison of multi-bit watermarking methods on accuracy (match rate). The match rate is the percentage of watermarked texts from which the full hidden message is correctly extracted. Texts were generated with \texttt{Llama~2~7B} and embedded message was 24 bits long.}
\label{tab:baselines}
\centering
\begin{tabular}{lcc}
\toprule
\textbf{Method} & \textbf{Match Rate} & \textbf{Limitation}\\
\midrule
Fernandez \emph{et al.}~\cite{fernandez2023} & $99.6\%$ & very long or infeasible extraction time \\
Wang \emph{et al.}~\cite{wang2024} & $98.8\%$ & very long or infeasible extraction time \\
Yoo \emph{et al.}~\cite{yoo2024} & $30.4\%$ & much below 98--100\% match rate \\
Lin \emph{et al.}~\cite{lin2024dermark} & 90.0\% & below 98--100\% match rate \\
Qu \emph{et al.}~\cite{qu2025} & $96.0\%$ & below 98-100\% match rate \\

\bottomrule
\end{tabular}
\end{table*}

On one end of the spectrum, \textit{enumeration-based} methods treat message extraction as a search problem, examining many possible bit sequences to find which one was embedded. Fernandez \emph{et al.}~\cite{fernandez2023} propose an enumeration approach that achieves very high extraction accuracy (nearly 99\% success on 16-bit payloads) by exhaustively searching for the hidden message. However, this comes at the cost of extremely slow decoding: extracting a 32-bit watermark from a single short text was reported to take on the order of 8~hours, making it impractical for large-scale deployment. Another multi-bit method by Wang \emph{et al.}~\cite{wang2024} improves upon this with a more efficient coding strategy, managing to extract 32-bit messages in about 2.3~hours per text in experiments. While faster, even a two-hour extraction for one short document is far too slow for real-time or mass detection scenarios. Moreover, both of these methods exhibit exponentially growing extraction time as the message length increases, due to their reliance on combinatorial search. This lack of scalability means that although they can reach high accuracy, they are not suitable for embedding more than a handful of bits in short texts.

At the other end of the design spectrum are approaches that prioritize efficiency by avoiding expensive search. A notable example is the method of Yoo \emph{et al.}~\cite{yoo2024}, which uses a position-based token allocation technique to encode bits on the fly during text generation. This approach enables nearly instantaneous extraction as it does not require iterating over candidate messages. However, the drawback is significantly lower accuracy in recovering the full message, especially as the payload size grows. For instance, Yoo et al.'s watermark could only recover a 16-bit message from a short text about 74\% of the time, dropping to around 30\% success for 24-bit payloads. Such error rates are far from acceptable for reliable tracing---losing even a single bit can corrupt the entire extracted identifier. In summary, purely efficient schemes tend to underperform on short texts, because a short generated passage provides limited opportunities to redundantly encode each bit and ensure error-free decoding.

A key observation from Table~\ref{tab:baselines} is that no existing method meets the $>$98\% extraction success criterion on short texts \textit{without} incurring prohibitive extraction time. The enumeration-based watermark of Fernandez \emph{et al.}~\cite{fernandez2023} achieves near-perfect accuracy on short passages, but only by tolerating an infeasible decoding latency. The fast method of Yoo \emph{et al.}~\cite{yoo2024} sacrifices too much accuracy to be useful for robust tracing. Wang \emph{et al.}'s~\cite{wang2024} balanced coding strikes a middle ground, but still does not reach the reliability needed for high-stakes use cases. Lin \emph{et al.}~\cite{lin2024dermark} propose \textsc{DERMARK}, which uses dynamic segment allocation to improve capacity on short texts; in roughly 100-token outputs it reports about 90\% success for 24-bit payloads, reducing tokens-per-bit compared to prior art, yet still falling short of the $>$98\% target in this regime. Very recently, Qu \emph{et al.}~\cite{qu2025} introduced a watermarking scheme that combines ideas from both camps---achieving 96--98\% extraction success on 20--24 bit messages in $\sim$200-token texts, while keeping extraction time around 1 second. This represents the state of the art to date, yet even this approach sees the success rate dip to 94\% for 32-bit payloads in a paragraph of text, and the performance on extremely short texts (e.g., single-sentence inputs) has not been fully demonstrated. In other words, the problem is not completely solved: there remains a practical gap in being able to robustly watermark and decode \textit{very short} AI-generated texts at scale.

As shown above, existing techniques each fail to fully satisfy the needs of short-text watermarking. An ideal solution would combine the near 100\% reliability of exhaustive methods with the speed of lightweight methods. This is precisely the challenge our work tackles.

\begin{figure*}[t]
  \centering
  \includegraphics[width=\textwidth]{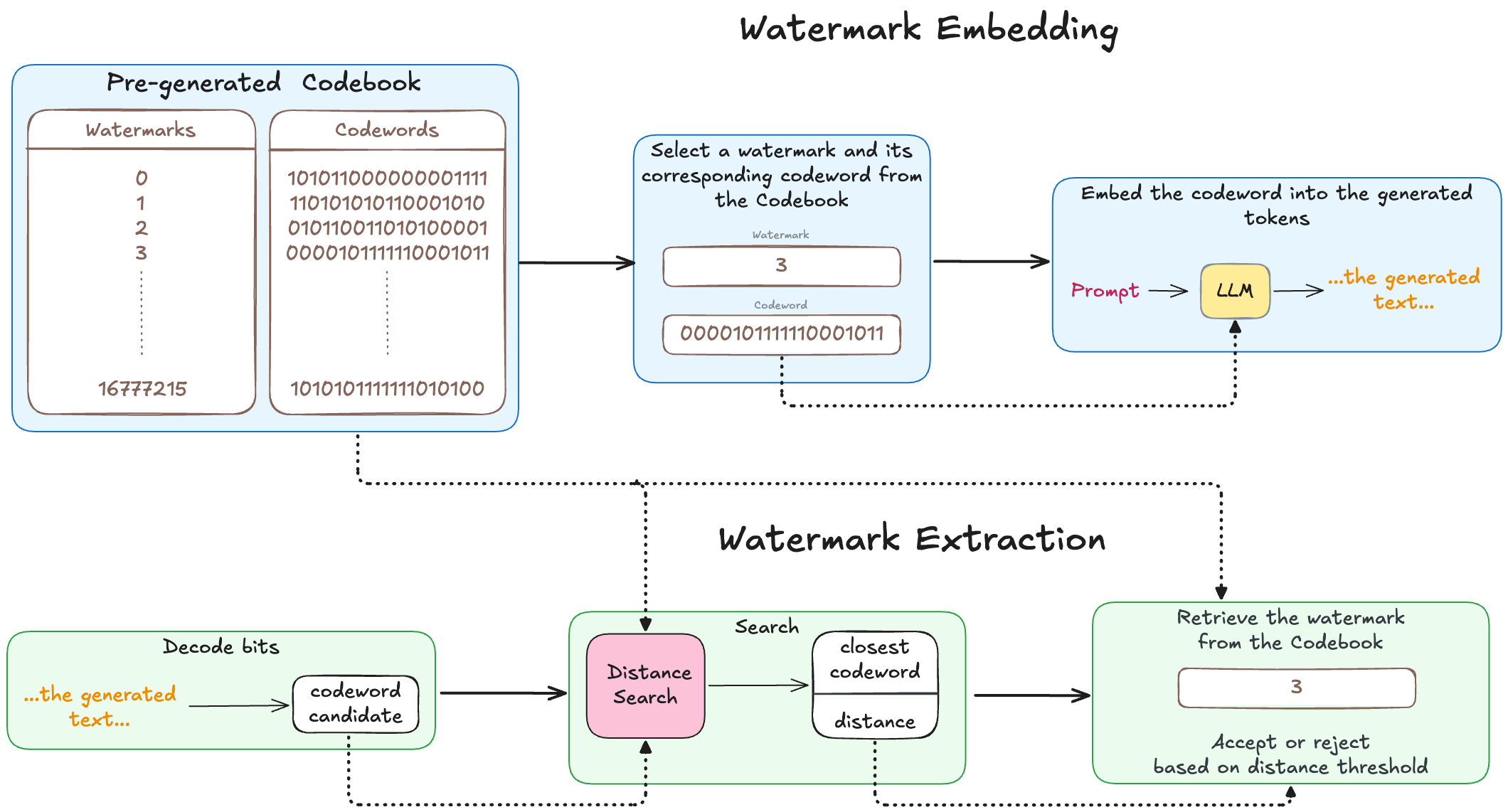}
  \caption{Our watermarking scheme. A codebook is pre-generated offline and later used by the embedding and extraction algorithms. The embedding phase embeds a long codeword (e.g., 127 bits) into the generated text. During the extraction, a noisy version of codeword is first obtained and then the closest codeword is searched in the codebook. Once the closest codeword is identified, the corresponding watermark is extracted. The detection decision (watermarked vs. non-watermarked) is based on the distance threshold. A distance is measured by the Hamming or Levenshtein distance.}
  \label{fig:watermarking_scheme}
\end{figure*}

\subsection{Contributions}
Our main contributions are as follows:
\begin{itemize}
    \item We developed a novel watermarking scheme with very high extraction success on multi-bit watermarks, even for short AI-generated texts, without sacrificing efficiency. We achieved a very high match rate with a 24-bit payload capacity, even on texts as short as 127 tokens in most settings. Our method also performs well at the standard 200-token evaluation length, matching or outperforming existing approaches.

    \item We showed that our main variant with Levenshtein distance provides a meaningful level of robustness and for some attacks is more robust than the state-of-the-art solutions.

    \item We investigated variants with Hamming and Levenshtein distances, analyzing their codeword collision and separation features. Additionally, we proposed the BCH variant, which does not require a pre-generated codebook. The BCH variant offers longer watermarks at the expense of lower accuracy and robustness.

\end{itemize}

\section{Multi-bit Watermarking Scheme with Codebook}
\label{sec:watermark_scheme}

To overcome the trade-off between accuracy and efficiency, we introduce a novel scheme centered around a pre-computed \textit{codebook}. Enumeration-based methods can be prohibitively slow not necessarily because they consider more candidate watermarks, but because evaluating each candidate may require expensive decoding. Our method still searches over all $2^n$ watermark identities, but it reduces extraction to a nearest-neighbor lookup against a precomputed codebook of $u$-bit codewords, where distance computations are very cheap and the codewords are well separated. This approach combines the robustness of having a long, redundant signal with the search efficiency, making it ideal for short texts where every token counts.

Our method consists of three stages: an offline Codebook Generation step, and online Watermark Embedding and Watermark Extraction steps.

\subsection{Codebook Generation}

\begin{algorithm*}[tbp]
\caption{Codebook Generation}
\label{alg:codebook_generation}
\begin{algorithmic}[1]
    \State \textbf{Input:} secret key $\mathcal{K}$, parameters $n,u \in \mathbb{N}$
    \State \textbf{Output:} codebook $\mathcal{C} : \{0,\dots,2^n-1\} \to \{0,1\}^u$

    \vspace{2mm}\hrule\vspace{2mm}

    \State $\mathbf{G}, \mathbf{r} \gets \mathsf{DeriveParams}(\mathcal{K}, n, u)$ \Comment{$n \times u$ binary matrix and $u$-bit offset (Appendix~\ref{sec:linear_codebook})}
    \State $\mathcal{C} \gets$ empty map
    \For{$i \gets 0$ to $2^n-1$}
       \State $W \gets i$
       \State $L \gets \mathrm{bits}(W) \cdot \mathbf{G} \bmod 2 \oplus \mathbf{r}$ \Comment{random linear code with affine offset}
       \State $\mathcal{C}[W] \gets L$
    \EndFor
    \State \Return $\mathcal{C}$
\end{algorithmic}
\end{algorithm*}

The first step, performed offline, is to generate a codebook $\mathcal{C}$ that serves as a shared secret between the embedding and extraction processes. As detailed in Algorithm \ref{alg:codebook_generation}, the codebook is a map from an $n$-bit watermark space to a $u$-bit codeword space, where $u > n$.

Specifically, we generate a table containing $2^n$ entries. Each entry maps a potential watermark $W$ (an integer from $0$ to $2^n-1$) to a unique binary vector $L_W$ of length $u$. To create this mapping, we use a random linear code over $\mathrm{GF}(2)$: we first derive a random $n \times u$ binary generator matrix $\mathbf{G}$ and a $u$-bit offset vector $\mathbf{r}$ from the secret key $\mathcal{K}$ (see Appendix~\ref{sec:linear_codebook} for details). Each codeword is then computed as $L_W = \mathrm{bits}(W) \cdot \mathbf{G} \bmod 2 \oplus \mathbf{r}$, where $\mathrm{bits}(W)$ is the $n$-bit binary representation of $W$. This process ensures that each message $W$ corresponds to a unique codeword, and without the secret key $\mathcal{K}$, the codebook cannot be reconstructed.

The redundancy introduced by making the codeword length $u$ greater than the message length $n$ is the key to our scheme's error resilience and effectiveness. The difference, $u-n$, can be seen as providing redundancy analogous to parity or check bits in classical error-correcting codes. By expanding each $n$-bit message into a $u$-bit codeword, we ensure that the valid codewords are, with high probability, far apart from each other in terms of the Hamming or Levenshtein distance. This separation allows us to correctly identify the original watermark even if several bits of the codewords are not encoded properly during generation. The greater the difference $u-n$, the more errors can be tolerated, improving the chances of perfect watermark recovery. However, the choice of these parameters presents a trade-off: choosing too small of an $n$ limits the total number of unique watermarks that can be embedded, while choosing too large of a $u$ requires longer generated texts.

The generation and storage of the codebook both have a time and memory complexity of $O(2^n \cdot u)$, as the process must iterate through all $2^n$ watermarks to generate a $u$-bit codeword for each. This is a one-time, offline cost. For the parameters used in our testing ($n=24$, $u=127$), the codebook requires storing $2^{24} \times 127$ bits, which amounts to a manageable 256~MB of memory. However, scaling to a 32-bit message space ($n=32$), which can support over 4 billion unique watermarks, would increase this requirement to $2^{32} \times 127$ bits, or a substantial 64~GB.

\subsection{Watermark Embedding}

The embedding process, shown in Algorithm \ref{alg:watermark_embedding}, encodes the watermark during the LLM's text generation phase. It modifies the token selection process to encode the bits of a chosen codeword $L_W$.

Given a watermark $W$ to embed, we first retrieve its corresponding $u$-bit codeword $L_W$ from the codebook $\mathcal{C}$. Then, for the first $u$ tokens to be generated, we embed bits from $L_W$. At each  token generation step $t$, we follow the standard practice from Kirchenbauer \emph{et al.}~\cite{kirchenbauer2023}, we use the secret key $\mathcal{K}$ and the previously generated token $s^{(t-1)}$ to deterministically partition the model's vocabulary $\mathcal{V}$ into two sets, which we call $\mathcal{V}_0$ and $\mathcal{V}_1$.

The $t$-th bit of the codeword, $b_t$, determines which vocabulary set is favoured. If $b_t=0$, we add a large positive bias $\delta$ to the logits of all tokens in $\mathcal{V}_0$. Likewise, if $b_t=1$, we bias the logits of tokens in $\mathcal{V}_1$. Having modified the logits, we use greedy sampling to select the next token $s^{(t)}$. The strong bias ($\delta=6$) makes it likely that the selected token $s^{(t)}$ will be a member of the favoured vocabulary set, therefore encoding the bit $b_t$. This procedure is repeated for all bits of the codeword, embedding the entire sequence $L_W$ into the tokens of the generated text $\mathcal{T}$. To mitigate tokenization inconsistency during embedding, we apply a stepwise consistency check (Appendix~\ref{sec:ti_correction}).

\begin{algorithm*}[tbp]
\caption{Watermark Embedding}
\label{alg:watermark_embedding}
\begin{algorithmic}[1]
    \State \textbf{Input:} large language model $LLM$, vocabulary $\mathcal{V}$, bias $\delta$, prompt $s^{(-N_p)},\dots,s^{(-1)}$, secret key $\mathcal{K}$, hash $\operatorname{hash}$, codebook $\mathcal{C}:\{0,\dots,2^n-1\} \to \{0,1\}^u$
    \State \textbf{Output:} generated tokens $\mathcal{T}$ with embedded watermark

    \vspace{2mm}\hrule\vspace{2mm}

    \State Choose a watermark $W \in \mathrm{domain}(\mathcal{C})$ \Comment{select a watermark from the codebook $\mathcal{C}$}
    \State $L_{W} \gets \mathcal{C}[W]$; \quad $u \gets |L_{W}|$; \quad \text{unpack } $L_{W}$ as $(b_0,\dots,b_{u-1})$
    \State $\mathcal{T} \gets [\,]$, \quad $t \gets 0$ \Comment{start embedding at the first generated token}

    \While{$t < u$}
        \State $seed \gets \begin{cases}
            \operatorname{hash}(\mathcal{K}), & \text{if } t=0 \\[1mm]
            \operatorname{hash}(s^{(t-1)}, \mathcal{K}), & \text{otherwise}
        \end{cases}$ \Comment{use $\operatorname{hash}(\mathcal{K})$ only for the very first token}
        \State Using $seed$, deterministically split $\mathcal{ V}$ into two equal sets $\mathcal{ V}_0$ and $\mathcal{ V}_1$
        \State $v \gets LLM(s^{(-N_p)},\dots,s^{(t-1)})$ \Comment{generate logits}

        \If{$b_t = 0$}
            \State $\forall\, w \in \mathcal{V}_0:\ \hat{v}[w] \gets v[w] + \delta$
        \Else
            \State $\forall\, w \in \mathcal{V}_1:\ \hat{v}[w] \gets v[w] + \delta$
        \EndIf

        \State $s^{(t)} \gets \operatorname{argmax}(\hat{v})$ \Comment{greedy sampling after bias}
        \State Append $s^{(t)}$ to $\mathcal{T}$; \quad $t \gets t+1$
    \EndWhile

    \State \Return $\mathcal{T}$
\end{algorithmic}
\end{algorithm*}

\subsection{Watermark Extraction}

The extraction process, detailed in Algorithm \ref{alg:watermark_extraction}, consists of two main phases: reconstructing the embedded codeword and finding the closest match in the codebook.

To reconstruct the embedded codeword, we iterate over the first $u$ tokens of the given text $\mathcal{T}$. For each token $s^{(t)}$ the vocabulary partitions $\mathcal{V}_0$ and $\mathcal{V}_1$ are regenerated based on the previous token $s^{(t-1)}$ and the secret key $\mathcal{K}$. We then check which group the token $s^{(t)}$ belongs to, if $s^{(t)} \in \mathcal{V}_1$ it is decoded as $1$, likewise if $s^{(t)} \in \mathcal{V}_0$ it is decoded as $0$. This decoding continues until a sequence of $u$ bits is recovered, which now acts as a codeword candidate $\widehat{L}$.

In the second phase, we find the closest match for this candidate codeword $\widehat{L}$ within the pre-computed codebook $\mathcal{C}$. We iterate through every watermark-codeword pair $(W, L_W)$ in the codebook and compute the distance, $d(\widehat{L}, L_W)$. Our main and recommended variant uses the Levenshtein distance but we also examine the Hamming distance. The watermark $W^\star$ corresponding to the codeword $L_{W^\star}$ that has the minimum distance to $\widehat{L}$ is selected as the most likely embedded message.

The extraction algorithm returns this best-match watermark $W^\star$ and its corresponding minimum distance, $d_{\min}$. This distance score serves as a confidence metric. A user can set a threshold $\eta$ for this distance. If $d_{\min} \leq \eta$, the watermark $W^\star$ is accepted. If the distance exceeds the threshold, the text can be classified as not containing a valid watermark or as having a signal that is too corrupted for reliable decoding. This error-correction capability stems directly from the redundancy in our codebook design ($u > n$). Because the valid codewords $L_W$ are generally some distance apart from each other in the Hamming and Levenshtein space, the system can tolerate a number of bit errors during generation (e.g., when low-entropy, peaky logits make the bias insufficient to steer selection into the intended partition) and still correctly identify the original watermark.

\begin{algorithm*}[tbp]
\caption{Watermark Extraction}
\label{alg:watermark_extraction}
\begin{algorithmic}[1]
    \State \textbf{Input:} tokens $\mathcal{T} = \big(s^{(0)}, s^{(1)}, \dots\big)$, vocabulary $\mathcal{V}$, secret key $\mathcal{K}$, hash $\operatorname{hash}$, codebook $\mathcal{C}$, codeword length $u$
    \State \textbf{Output:} $(W^\star, d_{\min})$ where $W^\star \in \mathrm{domain}(\mathcal{C})$ and $d_{\min} = d(\widehat{L},\, L_{W^\star})$

    \vspace{2mm}\hrule\vspace{2mm}

    \For{$t \gets 0$ to $u-1$}
        \State $seed \gets \begin{cases}
            \operatorname{hash}(\mathcal{K}), & \text{if } t=0 \\[1mm]
            \operatorname{hash}(s^{(t-1)}, \mathcal{K}), & \text{otherwise}
        \end{cases}$
        \State Using $seed$, deterministically split $\mathcal{V}$ into two equal sets $\mathcal{V}_0$ and $\mathcal{V}_1$
        \State $\widehat{b}_t \gets \begin{cases}
          1, & \text{if } s^{(t)} \in \mathcal{V}_1\\
          0, & \text{otherwise}
        \end{cases}$
    \EndFor

    \State $\widehat{L} \gets (\widehat{b}_0,\dots,\widehat{b}_{u-1})$
    \Statex \Comment{Search the stored codebook for the closest codeword}
    \State $(W^\star, d_{\min}) \gets (\bot, \infty)$

    \For{$(W, L_W)$ in $\mathcal{C}$}
        \State $d \gets d(\widehat{L}, L_W)$ \Comment{calculate the distance}
        \If{$d < d_{\min}$}
            \State $(W^\star, d_{\min}) \gets (W, d)$
        \EndIf
    \EndFor

    \State \Return $(W^\star, d_{\min})$ \Comment{a distance threshold may be applied as desired}
\end{algorithmic}
\end{algorithm*}

\subsection{Codeword Collision and Separation}
\label{sec:collision_separation}

Because each codeword is computed as $L_W = \mathrm{bits}(W) \cdot \mathbf{G} \bmod 2 \;\oplus\; \mathbf{r}$ (see~(\ref{eq:codeword})), two distinct watermarks $W_i \neq W_j$ collide if and only if $(\mathrm{bits}(i) \oplus \mathrm{bits}(j)) \cdot \mathbf{G} = \mathbf{0} \pmod{2}$. Since $\mathbf{G}$ is drawn uniformly at random, every nonzero linear combination of its rows is a uniformly random $u$-bit vector, so each pairwise collision occurs with probability $2^{-u}$. A union bound over all $\binom{2^n}{2} < 2^{2n-1}$ pairs gives:
\begin{equation}
    \Pr[\text{any collision}] \;\leq\; 2^{\,2n - u - 1}.
    \label{eq:collision_bound}
\end{equation}
For our parameters $n = 24$, $u = 127$, this evaluates to $2^{-80} \approx 10^{-24}$, making collisions negligible in practice.

Beyond collisions, the minimum Hamming distance $d_{\min}$ between any two codewords determines the scheme's error tolerance: nearest-neighbour decoding correctly identifies a watermark as long as at most $t = \lfloor (d_{\min} - 1) / 2 \rfloor$ bits are corrupted. For a random binary linear $[u, n]$ code, the Gilbert--Varshamov (GV) bound~\cite{macwilliams1977} guarantees the existence of a code with $d_{\min} \geq d_{\mathrm{GV}}$ satisfying $H_2(d_{\mathrm{GV}} / u) = 1 - n/u$, where $H_2$ is the binary entropy function. Evaluating this for our parameters yields $d_{\mathrm{GV}} \approx 32$ for $u = 127$ and $d_{\mathrm{GV}} \approx 60$ for $u = 200$. In our experiments (Appendix~\ref{sec:codebook_spread}), the measured minimum distances are $d_{\min} = 35$ and $d_{\min} = 62$, respectively---both exceeding the GV bound. This translates to an error tolerance of $t = 17$ bit errors for $u = 127$ and $t = 30$ for $u = 200$.

Finally, the linear construction offers a structural advantage over independent random (RNG) codewords. In a linear code, $d_{\min}$ is determined by the minimum Hamming weight among only $2^n - 1$ nonzero codewords, whereas RNG codes must take the minimum over all $\binom{2^n}{2} \approx 2^{2n}$ pairwise distances---a much larger set subject to a stronger birthday-type effect. At $n = 24$ and $u = 127$, this yields $d_{\min} = 35$ for the linear code versus a sampled minimum of~25 for RNG, a~40\% advantage. A detailed experimental comparison across all values of $n$ is provided in Appendix~\ref{sec:codebook_spread}.

\section{Variant with BCH Correction Code}
\label{sec:bch_correction_code}

The codebook-based watermarking scheme described in Section~\ref{sec:watermark_scheme} offers high accuracy and robustness for short texts, but it has notable limitations when scaling to larger watermark sizes. For example, with a 32-bit watermark (supporting over 4 billion unique identifiers), the codebook would require storing $2^{32}$ entries, which is computationally and memory-intensive during extraction. This makes the codebook approach impractical for very large watermark spaces, as discussed in Section~\ref{sec:limitations}.

To address this, we introduce a variant that uses a \textit{Bose--Chaudhuri--Hocquenghem} (BCH) error-correcting code instead of a codebook. The overall workflow remains similar: during encoding, a binary watermark is transformed into a binary codeword using BCH encoding, and this codeword is embedded into the text using the same token-level biasing strategy as in Algorithm~\ref{alg:watermark_embedding}. During extraction, the received codeword is decoded using the BCH algorithm to recover the original watermark. Unlike the codebook method, where the Hamming/Levenshtein distance to the closest codeword serves as a confidence metric, here the number of bit flips (i.e., the difference between the retrieved codeword and the codeword reconstructed from the decoded watermark) provides a measure of reliability. This approach avoids the need for a large codebook and enables efficient watermarking for larger payload sizes, at the cost of potentially lower error tolerance compared to the codebook method.

\section{Experimental Results}
\label{sec:results}

We compared our approach with the other top five methods. The results of the experiments are given in Table~\ref{tab:comparison-others}. Baseline numbers are taken from \cite{qu2025} and the experiment setting we followed is identical to the one used in that work. Watermarks of 16 to 32 bits are embedded in generated texts of 200 tokens. Texts were generated with \texttt{Llama~2~7B} and prompts were taken from the OpenGen dataset \cite{opengendataset}. All experiments used a consistent logit bias of $\delta=6$. We used greedy sampling, the same as \cite{qu2025} in their experiments.

The primary metric is \textit{match rate}, defined as the percentage of texts for which the full multi-bit watermark is correctly extracted.

Our codebook-based schemes achieve nearly a 100\% match rate across most settings, outperforming all other methods. For the 32-bit watermark, we obtain the second-best result (BCH variant), since the codebook-based algorithms were too computationally expensive (see Section~\ref{sec:limitations}).

We also evaluated our schemes in a more demanding setting with shorter 127-token texts embedding longer, 24-bit watermark. As shown in Table~\ref{tab:llmw_results}, the codebook methods achieve a high match rate across several models when using both Hamming and Levenshtein distance metrics. The table also compares against BCH variant. The codebook method consistently provides a higher match rate. The advantage is especially significant on instruction-tuned models, where the BCH scheme's performance degrades rapidly while our method remains robust. This highlights the superior error resilience of our pseudorandom codebook approach over the structured constraints of BCH codes, particularly in the challenging low-entropy settings of aligned models.

Generally, performance is comparatively lower on instruction-tuned models. We attribute this to the low-entropy, peaky logit distributions produced by these models after fine-tuning with methods like RLHF, which can overpower the logit-biasing mechanism.

\begin{table}[t]
\caption{Comparison of methods on match rate (\%) for different watermark lengths (in bits) on 200 tokens. Evaluated on the OpenGen dataset~\cite{opengendataset}. \textsuperscript{$\dagger$}BCH uses 127 tokens (native BCH codeword length) for 16--24-bit watermarks to maximize the error correction rate; the 32-bit watermark uses 200 tokens with code shortening (see Appendix~\ref{sec:bch_token_length}). NA denotes that a method was too computationally expensive for the whole dataset.}
\label{tab:comparison-others}
\centering
\setlength{\tabcolsep}{4pt}
\begin{tabular}{l|cccc}
    \toprule
    Watermark size &16&20&24&32 \\
    \midrule
    Fernandez \emph{et al.}~\cite{fernandez2023}    & 99.6          & 99.2          & 98.0          & NA \\
    Wang \emph{et al.}~\cite{wang2024}              & 98.8          & 98.4          & 97.2          & NA \\
    Yoo \emph{et al.}~\cite{yoo2024}                & 73.6          & 49.2          & 30.4          & 8.4 \\
    Cohen \emph{et al.}~\cite{Cohen2025WatermarkingAdaptiveUsers}   & 88.8          & 78.4          & 65.6          & 27.2 \\
    Qu \emph{et al.}~\cite{qu2025}                  & 98.0          & 97.6          & 96.0          & \textbf{94.0} \\
    Ours (BCH\textsuperscript{$\dagger$})       & 99.1          & 98.6          & 97.6          & 91.1 \\
    Ours (Hamming)                              & \textbf{99.9} & \textbf{99.8} & \textbf{99.9} & NA \\
    Ours (Levenshtein)                          & 99.8          & \textbf{99.8} & 99.8          & NA \\
    \bottomrule
\end{tabular}
\end{table}

\begin{table}[t]
\caption{Match rate (\%) at FPR $\leq 1\%$ for a 24-bit watermark. Evaluated on Essays-With-Instructions dataset~\cite{Schuhmann2023EssaysWithInstructions} and HC3 dataset~\cite{guo-etal-2023-hc3}.}
\label{tab:llmw_results}
\centering
\begin{tabular}{l|ccc}
    \toprule
    \textbf{Model} & \textbf{Hamming} & \textbf{Levenshtein} & \textbf{BCH} \\
    \midrule
    \multicolumn{4}{l}{\quad\textit{127 tokens}} \\
    \midrule
    \texttt{Llama~2~7B}              & 99.3  & 97.5  & 98.3  \\
    \texttt{Llama~3.1~8B}            & 100   & 100   & 99.7  \\
    \texttt{Llama~3.2~3B}            & 100   & 100   & 99.6  \\
    \texttt{Llama~3.2~3B~Inst.}      & 90.2  & 61.6  & 56.5  \\
    \texttt{Mistral~7B}              & 99.7  & 99.4  & 99.4  \\
    \texttt{Mistral~7B~Inst.}        & 96.3  & 76.4  & 64.5  \\
    \midrule
    \multicolumn{4}{l}{\quad\textit{200 tokens}} \\
    \midrule
    \texttt{Llama~2~7B}              & 100   & 99.6  & 99.3  \\
    \texttt{Llama~3.1~8B}            & 99.6  & 99.6  & 99.6  \\
    \texttt{Llama~3.2~3B}            & 100   & 100   & 99.6  \\
    \texttt{Llama~3.2~3B~Inst.}      & 99.6  & 90.6  & 39.6  \\
    \texttt{Mistral~7B}              & 100   & 99.7  & 98.7  \\
    \texttt{Mistral~7B~Inst.}        & 100   & 96.7  & 43.8  \\
    \bottomrule
\end{tabular}
\end{table}

\subsection{Extraction Time}
Extraction consists of decoding the bit sequence from text and searching the codebook for the closest match. Decoding is negligible (tokenization and per-token classification); the bottleneck is the brute-force search over all $2^n$ entries. The BCH variant bypasses this search entirely via algebraic decoding. Table~\ref{tab:extraction_time} reports the extraction time for $n{=}24$ (${\approx}16.8$M entries), averaged over 10{,}000 queries. The Hamming variant uses the Faiss library on CPU with 8 OpenMP threads; the Levenshtein variant uses a custom Triton kernel on the GPU.

\begin{table}[ht]
\caption{Average extraction time per query for $n{=}24$ (${\approx}16.8$M codebook entries). Hamming and Levenshtein benchmarked on an AMD Epyc 7302 (16C/32T, 3.0\,GHz) with a single Nvidia RTX~5090 (32\,GB).}
\label{tab:extraction_time}
\centering
\begin{tabular}{l|cc}
    \toprule
    \textbf{Variant} & $u{=}127$ & $u{=}200$ \\
    \midrule
    Levenshtein (GPU, Triton)     & 5.8\,ms & 75.5\,ms \\
    Hamming (CPU, Faiss)          & 4.1\,ms & 13.3\,ms \\
    BCH (CPU, algebraic)          & 2.3\,ms & 3.5\,ms \\
    \bottomrule
\end{tabular}
\end{table}

Even in the slowest setting, extraction takes under 0.1\,seconds---practical for real-time use. For comparison, other methods that reach comparable match rates are orders of magnitude slower for watermarks upwards of 20 bits~\cite{qu2025}.

\subsection{Quality of Watermarked Texts}
Our quality evaluation (Appendix~\ref{sec:text_quality}) indicates that the watermark has only a modest effect on perceived text quality. In an LLM-as-a-Judge study on review-style and social-post prompts, watermarked outputs were rated similarly to non-watermarked ones and even received a small authenticity preference. Perplexity increases under watermarking, but the values remain moderate. Lexical-diversity analysis further shows that watermarked text is consistently more diverse and less repetitive, suggesting that the bias introduces phrasing variation rather than obvious degradation. Overall, these results support that the watermark preserves readable, socially plausible text while slightly increasing diversity.

\subsection{Attack Robustness}
\label{sec:attack_robustness}

\begin{figure*}[t]
  \centering
  \includegraphics[width=\textwidth]{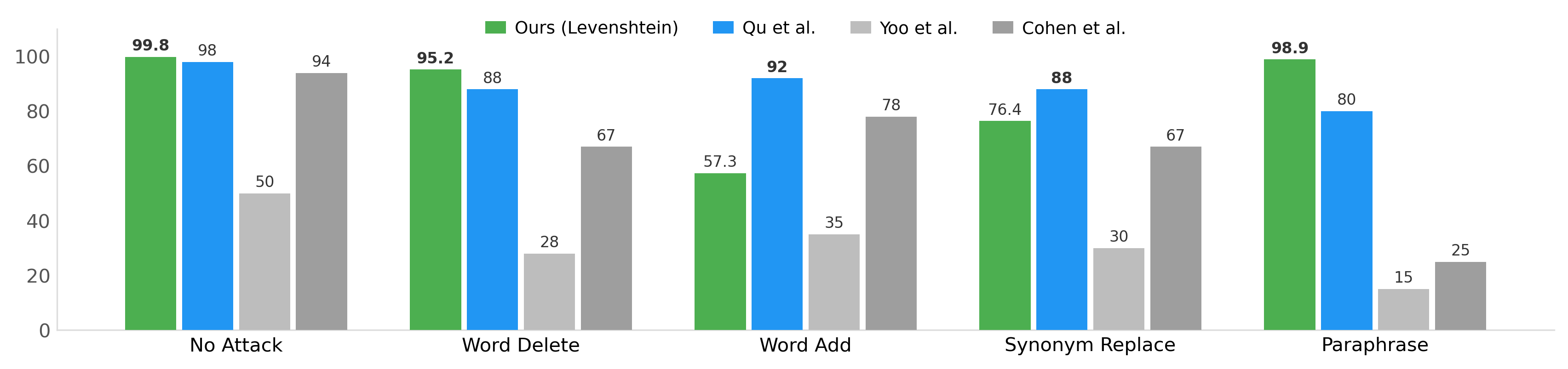}
  \caption{Match rate under attacks, evaluated on generations from \texttt{Llama~2~7B} with the OpenGen dataset~\cite{opengendataset}. The attacker's budget is 10\% of a total number of tokens for a given text. Baseline values are estimated from Figure~6 of Qu \emph{et al.}~\cite{qu2025}.}
  \label{fig:attack_robustness}
\end{figure*}

We consider an attacker whose aim is to remove the watermark and alter the multi-bit payload. The attacker can edit the watermarked text while preserving the semantics and readability.

Because we focus on short texts generated at scale, we exclude attacks that require substantial human effort (e.g., manual rewriting) or heavy paraphrasing by other models. In our regime, a strong global rewrite often replaces most of the surface form and becomes very close to generating a new text with the same intent. An attacker capable of doing this can often bypass watermarking by generating a fresh unwatermarked text directly.

We apply four common attacks:
word insertion, word deletion, substitution with synonyms and paraphrasing (details in Appendix~\ref{sec:attack_setup}). Following a common setting, we modify 10\% tokens. Figure~\ref{fig:attack_robustness} shows a comparison between our recommended variant (Levenshtein) and baselines \cite{yoo2024,Cohen2025WatermarkingAdaptiveUsers,qu2025}.

Our method offers a meaningful level of robustness, particularly against the word deletion and light paraphrasing attacks (10\% modified tokens). The word insertion attack is more challenging because it increases the Levenshtein distance to the correct codeword and the extraction algorithm recovers an incorrect watermark more often. Yet, it still achieves nearly 60\% match rate.

Hamming and BCH variants are brittle against the considered attacks. This is because they compare codewords bit-by-bit at fixed positions without considering positional shifts. Even a single insertion or deletion causes a frameshift that desynchronizes all subsequent bit comparisons. Therefore, for applications where robustness matters, we recommend a scheme which uses the Levenshtein distance.

\section{Limitations}
\label{sec:limitations}
A main limitation of our proposed scheme is the size of the codebook, which directly impacts extraction efficiency. While the watermark space, determined by $n$, can theoretically be made large (e.g., $n=32$ to support over 4 billion unique watermarks), this creates a practical bottleneck during the extraction phase. The extraction process requires computing a distance between the candidate codeword $\widehat{L}$ and every codeword $L_W$ in the codebook $\mathcal{C}$. For a large $n$, this computation becomes prohibitively expensive, undermining the goal of a fast and scalable detection system. Therefore, a trade-off exists between the watermark's capacity ($2^n$) and the computational cost of extraction.

The limitation can be circumvented by requiring longer texts for embedding a watermark. For example, instead of common 200-token text, we would use 256 tokens. Then, we can easily embed 32-bit watermark by dividing the text into two parts and embedding 16-bit watermark in each part.

\section{Conclusion}
\label{sec:conclusions}
We proposed a novel multi-bit watermarking scheme for LLM-generated texts. We showed that for watermark lengths up to 24 bits, our approach can achieve both near-100\% match rates and practical extraction speed. For most models, this stays true even for very short, 127-token texts. This makes the method well suited to watermarking short comments, product reviews or social media posts --- all potential AI-generated fake content. Our recommended Levenshtein variant offers solid robustness against common attacks. We also investigated other variants that may be preferable when speed is the primary concern (Hamming) or when longer watermarks are required (BCH).

\appendices

\section{AUROC}
\label{sec:auroc}

\begin{figure}[ht]
  \centering
  \includegraphics[width=\columnwidth]{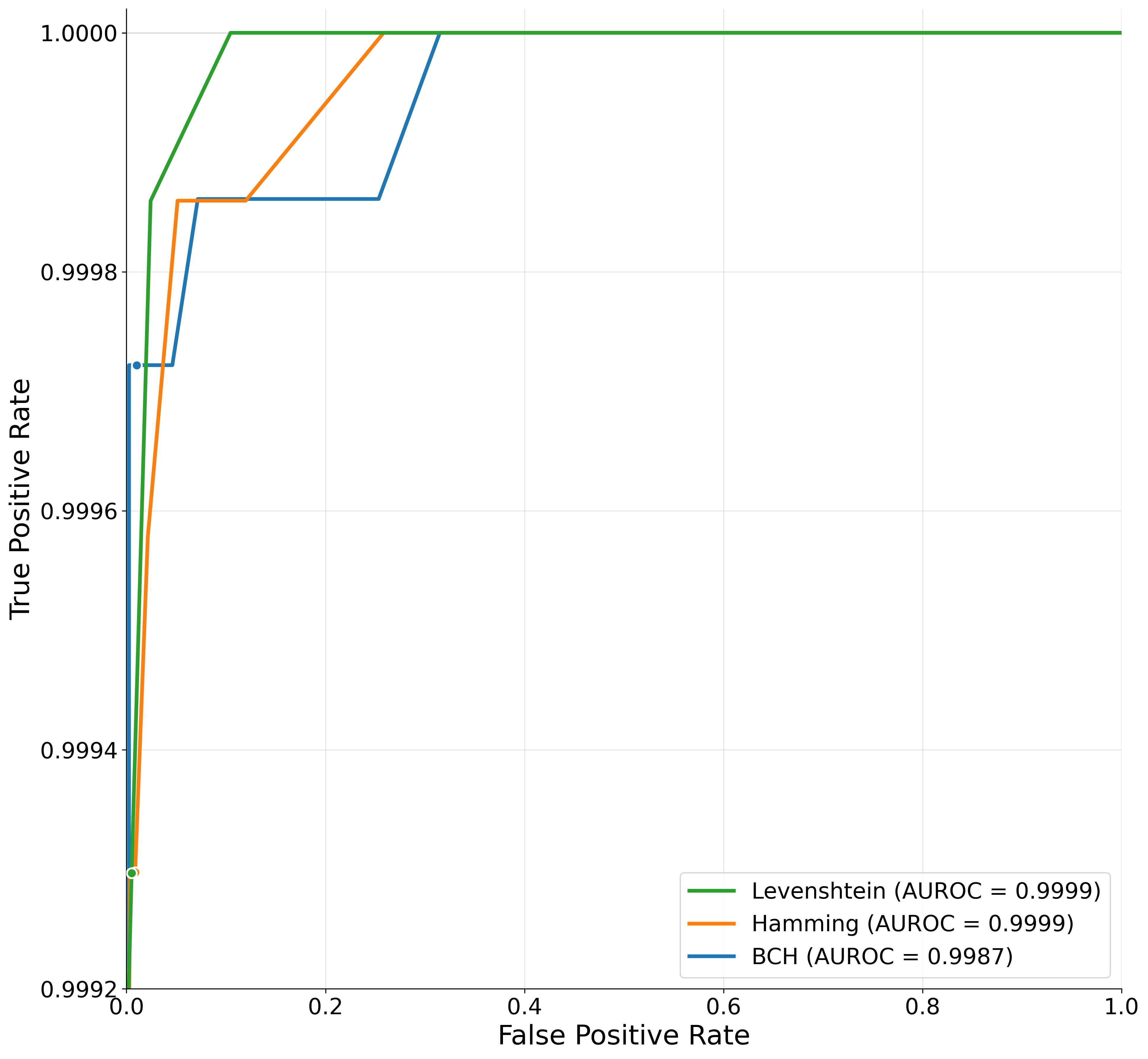}
  \caption{ROC curve with reported AUROC. Marked points have 1\% False Positive Rate.}
  \label{fig:roc_curve}
\end{figure}

Figure~\ref{fig:roc_curve} shows the combined ROC curves for our three watermarking variants. All three methods achieve near-perfect discrimination between watermarked and non-watermarked text, with AUROC values above 0.998. This evaluation was performed on \texttt{Llama~2~7B} generations from the OpenGen dataset~\cite{opengendataset} with a 24-bit watermark ($n=24$). The Hamming and Levenshtein variants are evaluated on 200 tokens; the BCH variant uses $u=127$ tokens, its native BCH codeword length, to maximize the error correction rate (see Appendix~\ref{sec:bch_token_length}).

\section{Attack Robustness Setup}
\label{sec:attack_setup}

The attack robustness experiments in Section~\ref{sec:attack_robustness} apply adversarial modifications at the \emph{word level} to the watermarked text itself. This reflects a realistic threat model in which an attacker edits the generated text while attempting to preserve its meaning and readability.

All four attacks are governed by an \textit{attack budget}, a fraction of tokens to be affected. Given a watermarked text of length $T$~tokens (measured by the model's tokenizer without special tokens), the target number of tokens to modify is $B = \lfloor \text{attack\_budget} \cdot T \rfloor$. In our experiments we set the attack budget to 0.1 (10\% of tokens).

\subsubsection{Word deletion}
The text is split into words by whitespace. The word indices are shuffled uniformly at random and words are deleted one by one, accumulating the token count of each removed word, until the cumulative count reaches or exceeds~$B$. Because tokens are removed, the decoded bit sequence becomes shorter than~$u$, causing positional shifts in the remaining bits.

\subsubsection{Word insertion}
Random English words are drawn from the NLTK words corpus (approximately 235{,}000 entries) and inserted at uniformly random positions in the text until the cumulative token count of inserted words reaches~$B$. Inserted tokens lengthen the decoded bit sequence beyond~$u$ and shift all subsequent bit positions, making this attack particularly challenging for fixed-position distance metrics.

\subsubsection{Synonym replacement}
Words are part-of-speech tagged using the Penn Treebank tagset and mapped to WordNet categories (noun, verb, adjective, adverb). For each word with at least one single-word synonym in WordNet, we record it as a replacement candidate. Candidates are shuffled and replaced one by one with a randomly chosen synonym until the cumulative token count of the original words reaches~$B$. Unlike deletion and insertion, synonym replacement largely preserves text length, so its primary effect is to alter the decoded bit values at the affected positions without shifting neighbouring bits.

\subsubsection{Paraphrasing}
A contiguous region of words is selected by choosing a random centre position and expanding alternately left and right until the region spans at least~$B$~tokens. The selected region is sent to a large language model (\texttt{GPT-oss-120B} via an OpenAI-compatible API) with the instruction: \textit{``Rewrite the following text while preserving its meaning. Output only the rewritten text, nothing else.''} ($\texttt{temperature}=0.7$). The paraphrased region replaces the original, while the surrounding prefix and suffix remain unchanged. This attack can simultaneously alter bit values, insert new tokens, and delete existing ones within the affected region.

\subsubsection{Evaluation procedure}
After each attack, the modified text is re-tokenized and decoded using the same green/red token classification as during embedding (Section~\ref{sec:watermark_scheme}). If the decoded bit sequence exceeds~$u$ bits (which can occur after word insertion, synonym replacement, or paraphrasing), it is truncated to exactly~$u$ bits. The resulting bit sequence is searched against the codebook using the Levenshtein distance, and the watermark identity of the closest codeword is returned. We report the \textit{match rate}: the fraction of samples for which the recovered watermark equals the originally embedded one.

\subsubsection{Experimental parameters}
The experiments in Figure~\ref{fig:attack_robustness} use watermarked texts generated by \texttt{Llama~2~7B} on the OpenGen dataset~\cite{opengendataset}, with parameters $n = 20$, $u = 200$, logit bias $\delta = 6$, and a linear codebook searched with the Levenshtein distance. The attack budget is 0.1.
(These are the same attack settings as in Qu \emph{et al.}~\cite{qu2025}.)

\section{Codebook Construction}
\label{sec:linear_codebook}

The $\mathsf{DeriveParams}$ procedure in Algorithm~\ref{alg:codebook_generation} deterministically derives the generator matrix $\mathbf{G} \in \{0,1\}^{n \times u}$ and offset vector $\mathbf{r} \in \{0,1\}^{u}$ from the secret key $\mathcal{K}$ as follows.

Two independent seeds are obtained by hashing the key with distinct domain-separation tags:
\begin{align}
    \mathit{seed}_{\mathbf{G}} &= \mathrm{SHA256}\!\left(\mathcal{K} \,\|\, \texttt{"linear-G"}\right)\![{:}64] \label{eq:seed_g} \\
    \mathit{seed}_{\mathbf{r}} &= \mathrm{SHA256}\!\left(\mathcal{K} \,\|\, \texttt{"linear-r"}\right)\![{:}64] \label{eq:seed_r}
\end{align}
where $\|$ denotes string concatenation and $[{:}64]$ selects the first 64 bits of the digest. Each seed initializes a pseudorandom number generator that fills its respective structure with i.i.d.\ $\mathrm{Bernoulli}(0.5)$ bits: $\mathit{seed}_{\mathbf{G}}$ produces all $n \cdot u$ entries of $\mathbf{G}$, and $\mathit{seed}_{\mathbf{r}}$ produces the $u$ entries of $\mathbf{r}$.

Given $\mathbf{G}$ and $\mathbf{r}$, the codeword for watermark $W$ is:
\begin{equation}
    L_W = \mathrm{bits}(W) \cdot \mathbf{G} \bmod 2 \;\oplus\; \mathbf{r}
    \label{eq:codeword}
\end{equation}
where $\mathrm{bits}(W) \in \{0,1\}^{n}$ is the big-endian binary representation of $W$. The set of all codewords forms a coset of a random binary linear code, shifted by the offset~$\mathbf{r}$.

An alternative construction generates codewords independently at random. For each watermark identity~$i$, a seed is derived as $\mathrm{SHA256}(\mathcal{K} \,\|\, i)$ and used to initialize a PRNG that samples $u$~i.i.d.\ $\mathrm{Bernoulli}(0.5)$ bits. We refer to this as the \textit{RNG} method. Because each codeword is drawn independently, the pairwise distances between codewords are random variables with no algebraic structure linking them.

\section{BCH Token Length Selection}
\label{sec:bch_token_length}

In Table~\ref{tab:comparison-others}, the BCH variant uses only 127~tokens for 16--24-bit watermarks, seemingly giving up the advantage of the extra 73~bits available to all other methods. This section explains why: BCH codes have fixed native codeword lengths of the form $2^m - 1$, and adapting them to non-native lengths requires code shortening that reduces the relative error correction rate, making the longer codeword counterproductively less robust.

\subsubsection{Native BCH lengths and code shortening}
A BCH code of field degree~$m$ has a native codeword length $n_0 = 2^m - 1$. For our experiments the relevant values are $n_0 = 127$ ($m = 7$) and $n_0 = 255$ ($m = 8$). When the desired output length~$u$ does not equal~$n_0$, the standard approach is \emph{code shortening}: a BCH code with $n_0 > u$ is used, but $n_0 - u$ message positions are fixed to known zeros and omitted from the transmitted codeword. Shortening preserves the absolute correction capability~$t$, but because the codeword shrinks while $t$ stays fixed, the relative correction rate $t/u$ decreases compared to using the code at its native length.

\subsubsection{Parameter comparison}
At $u = 127$ the BCH code operates at its native length ($n_0 = 127$)---a perfect fit requiring no shortening. At $u = 200$, the nearest larger native length is $n_0 = 255$, requiring 55~bits of shortening. Table~\ref{tab:bch_params} shows the resulting parameters for $n = 24$: although $t$ rises from 21 to 27 in absolute terms, the correction rate $t/u$ \emph{drops} from 16.5\% to 13.5\%.

\begin{table}[ht]
\caption{BCH parameters for a 24-bit watermark ($n = 24$) at two codeword lengths. At $u = 127$ the code operates at its native length with no shortening, yielding a higher relative error correction rate.}
\label{tab:bch_params}
\centering
\begin{tabular}{l|cc}
    \toprule
    & $u = 127$ & $u = 200$ \\
    \midrule
    Native length $n_0$ & 127 & 255 \\
    Field degree $m$ & 7 & 8 \\
    Message capacity $k$ & 29 & 79 \\
    Shortening & 0 & 55 bits \\
    Correction capability $t$ & 21 & 27 \\
    Correction rate $t / u$ & 16.5\% & 13.5\% \\
    \bottomrule
\end{tabular}
\end{table}

The same pattern holds for $n = 16$ and $n = 20$: the native $u = 127$ configuration yields correction rates of 18.1\% compared to 14.5\% and 13.5\%, respectively, at $u = 200$.

\subsubsection{Empirical evidence}
The instruction-tuned models in Table~\ref{tab:llmw_results} showcase this effect. These models produce low-entropy logit distributions that resist the embedding bias, pushing per-bit error rates above~14\%. At $u = 127$ the expected number of bit errors (${\approx}\,20$) falls just within the correction capability ($t = 21$), whereas at $u = 200$ the expected errors (${\approx}\,29$) exceed the limit ($t = 27$). Concretely, for $n = 24$, \texttt{Llama~3.2~3B~Instruct} achieves a 56.5\% match rate at $u = 127$ but only 39.6\% at $u = 200$; \texttt{Mistral~7B~Instruct} shows a similar gap of 64.5\% versus 43.8\%. On base models, where per-bit error rates are lower (${\approx}\,6$\%), both configurations succeed comfortably.

\subsubsection{Why $n = 32$ uses $u = 200$}
For a 32-bit message at $u = 127$, the BCH code requires $k = 36$, leaving fewer parity bits and reducing correction to $t = 15$ ($t/u = 11.8$\%). At $u = 200$, the code uses a BCH(255,\,87) code with shortening: the 32-bit watermark occupies the first 32 of the 87 message positions, the remaining~55 are set to known zeros, the full 255-bit codeword is encoded, and the 55~known-zero positions are removed to produce a 200-bit transmitted sequence. This yields $t = 26$ ($t/u = 13.0$\%), reversing the advantage: for $n = 32$, the shortened 200-bit code provides both better error resilience and a direct comparison with the other methods in Table~\ref{tab:comparison-others}, which all use 200~tokens.

\section{Codebook Spread and Separation}
\label{sec:codebook_spread}

To characterise how tightly packed the codebook becomes as the number of watermark identities grows, we measure the \textit{spread}: the nearest-neighbour distance distribution across sampled codewords. For each value of~$n$ from~4 to~24, we sample $z = 50{,}000$ codewords (or all $2^n$ if fewer exist) from the codebook and compute each codeword's distance to its nearest neighbour, excluding itself. We then report the minimum, mean, and maximum of these nearest-neighbour distances. As $n$~increases and the codebook grows denser in the binary space, we expect nearest-neighbour distances to decrease; the rate of this decrease and the spread between minimum and maximum reveal how uniformly the code distributes its codewords.

\subsubsection{Results}
\begin{figure}[ht]
  \centering
  \includegraphics[width=\columnwidth]{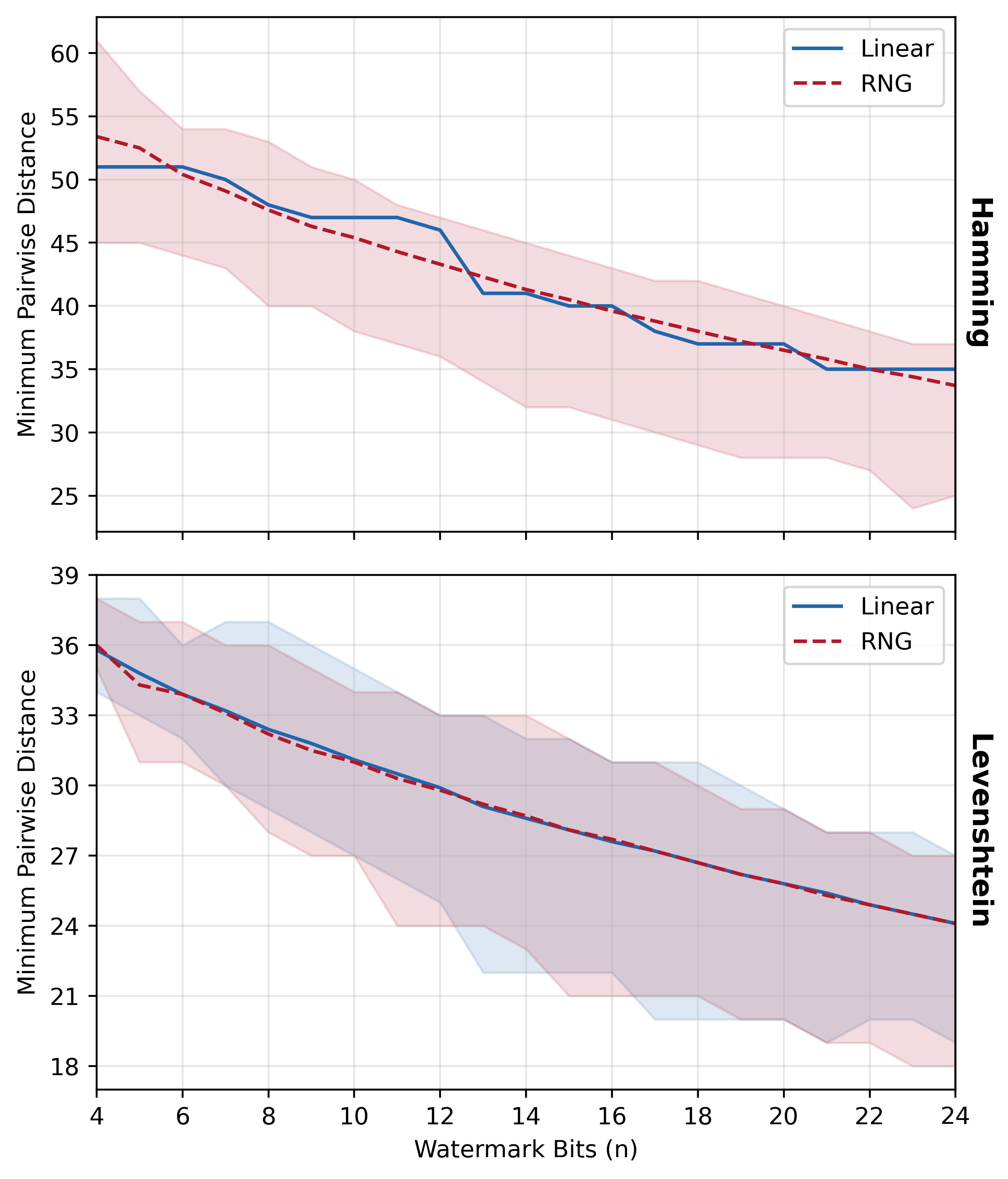}
  \caption{Nearest-neighbour distance statistics for Linear and RNG codebook construction with codeword length $u = 127$. Lines show the mean; shaded regions show the min--max range. The Linear method has no visible band under Hamming distance because min $=$ mean $=$ max at every~$n$.}
  \label{fig:spread_u127}
\end{figure}

\begin{figure}[ht]
  \centering
  \includegraphics[width=\columnwidth]{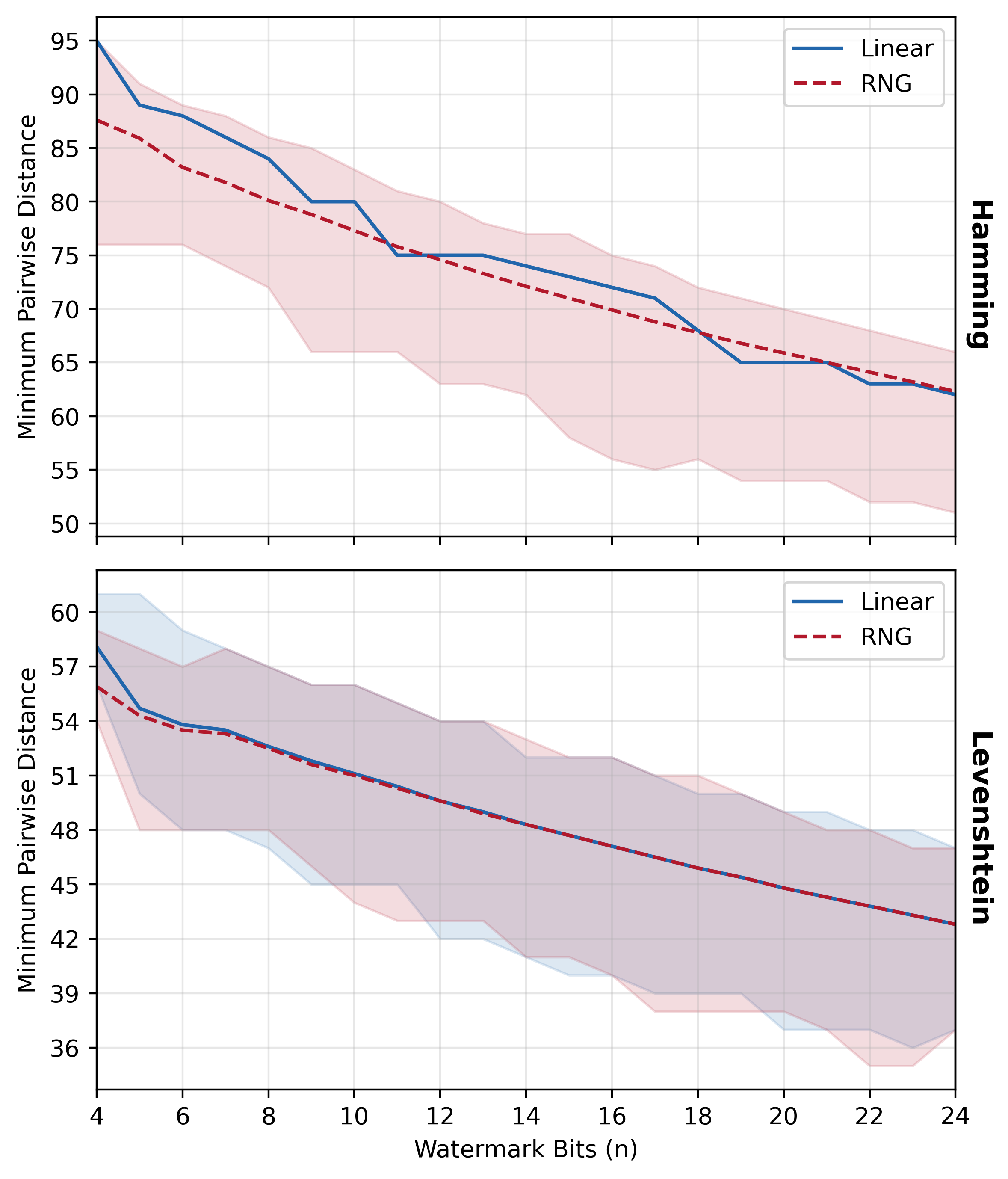}
  \caption{Same as Figure~\ref{fig:spread_u127} but for codeword length $u = 200$.}
  \label{fig:spread_u200}
\end{figure}

Figures~\ref{fig:spread_u127} and~\ref{fig:spread_u200} compare the Linear and RNG construction methods under both Hamming and Levenshtein distance for codeword lengths $u = 127$ and $u = 200$, respectively. Under Levenshtein distance, both methods behave nearly identically: the mean curves overlap and the min--max bands are comparable. Under Hamming distance, however, a striking difference emerges. The Linear construction produces \textit{zero spread}: the minimum, mean, and maximum nearest-neighbour distances are all equal at every~$n$, so no shaded band appears. The RNG method, by contrast, shows substantial spread, with the gap between minimum and maximum widening as~$n$ grows.

\subsubsection{Why Linear codes have zero Hamming spread}
This is a fundamental property of linear codes. For any two codewords $L_i$ and $L_j$:

\begin{equation*}
    L_i \oplus L_j = \bigl(\mathrm{bits}(i) \oplus \mathrm{bits}(j)\bigr) \cdot \mathbf{G} \bmod 2
\end{equation*}

since the offset~$\mathbf{r}$ cancels under XOR. The Hamming distance between $L_i$ and~$L_j$ therefore equals the Hamming weight of the corresponding codeword in the underlying linear code. A well-known property of linear codes is that the minimum distance equals the minimum Hamming weight of any non-zero codeword. Because every codeword in the code can appear as a difference $L_i \oplus L_j$ for some pair, the nearest neighbour of \textit{every} entry achieves this minimum weight. Hence the nearest-neighbour distance is identical for all codewords.

\subsubsection{Why Levenshtein distance breaks this uniformity}
Levenshtein (edit) distance permits insertions and deletions, which shift bit positions and break the algebraic structure that Hamming distance preserves. Two codeword pairs with the same Hamming distance can have different edit distances depending on \textit{where} the differing bits are located --- clustered differences can be bridged by fewer edit operations than scattered ones. Since the linear code's generator matrix~$\mathbf{G}$ has no control over the spatial distribution of differing bits, the nearest-neighbour edit distances vary across codewords even though their Hamming-distance neighbours are all equidistant.

\subsubsection{Choice of construction}
We adopt linear codebook construction for its deterministic guarantee of uniform minimum distance in Hamming space. Every codeword enjoys the same nearest-neighbour distance, providing uniform error tolerance across all watermark identities. The RNG method, while producing comparable mean distances, cannot offer this guarantee --- some identities may be unluckily close to a neighbour, reducing robustness for those watermarks.

\section{Tokenization Inconsistency Correction}
\label{sec:ti_correction}

Subword tokenizers used by modern LLMs are not round-trip stable: decoding a token sequence into text and re-encoding that text can produce a token list of different length~\cite{yan2025ti}. This phenomenon, known as \textit{tokenization inconsistency} (TI), arises because the greedy merge rules applied during encoding may segment a string differently when surrounding context changes. Yan and Murawaki~\cite{yan2025ti} report that while token-level inconsistency rates are typically below 0.3\% for \texttt{Llama~2~7B}, the text-level rate---the probability that \textit{at least one} token in a generated text is affected---reaches 33.6\% for 200-token sequences.

\subsubsection{Impact on our scheme}
In our watermarking pipeline, the extractor receives text rather than the original token IDs and must retokenize it to recover the embedded codeword. Because each token encodes exactly one bit, a single token insertion or deletion during retokenization shifts all subsequent bit positions. For Hamming distance and BCH decoding, which rely on strict positional alignment, this causes complete desynchronization of the extracted codeword, typically resulting in extraction failure. Levenshtein distance is only marginally affected, as it natively handles insertions and deletions, but the extra edit operations still reduce the effective codeword separation.

\subsubsection{Stepwise consistency verification}
To eliminate TI at its source, we augment the embedding procedure (Algorithm~\ref{alg:watermark_embedding}) with a per-token consistency check. After sampling each token $s^{(t)}$, we verify that the generated sequence $(s^{(0)}, \dots, s^{(t)})$ survives a decode--encode round-trip with the same length. If the check fails, $s^{(t)}$ is excluded by setting its logit to $-\infty$, and a new token is sampled from the remaining distribution. This process repeats until a consistent token is found or a retry limit is reached.

The key properties of this approach are:
\begin{itemize}
    \item \textbf{No additional model calls.} The model's logits are computed once per token position; retries only mask and resample from the same logit vector.

    \item \textbf{Length-only check.} Verifying that the round-trip preserves the sequence length is sufficient, because length preservation guarantees correct bit-position alignment during extraction.

    \item \textbf{Negligible overhead.} Each consistency check requires one tokenizer decode and one encode call. On \texttt{Llama~2~7B} the cumulative check time over a 200-token generation averages 139\,ms (0.70\,ms per token), roughly 1.2\% of the total generation time.
\end{itemize}

Our method is similar in spirit to the TI mitigation methods proposed by Yan and Murawaki~\cite{yan2025ti}. Since inconsistent tokens are infrequent resampling rarely occurs and the impact on text quality is negligible.

\section{Text Quality Evaluation}
\label{sec:text_quality}

\begin{figure*}[ht]
  \centering
  \includegraphics[width=\textwidth]{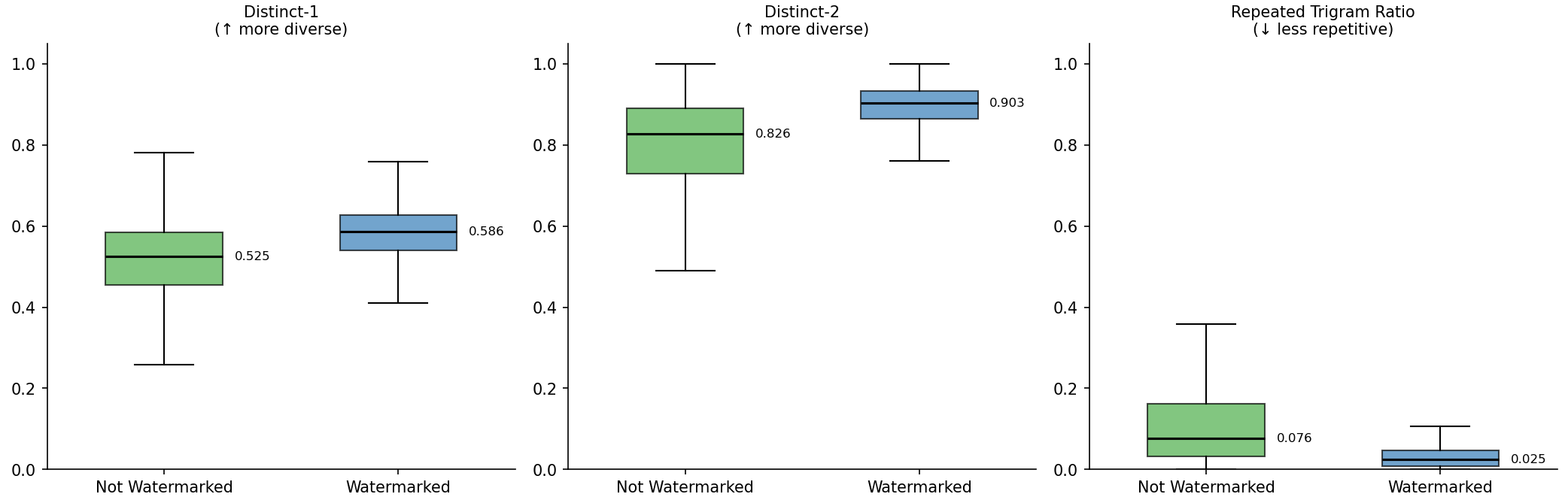}
  \caption{Lexical diversity distributions for 200-token \texttt{Llama~2~7B} generations on the OpenGen dataset~\cite{opengendataset}.}
  \label{fig:diversity_llama2}
\end{figure*}

\begin{figure*}[ht]
  \centering
  \includegraphics[width=\textwidth]{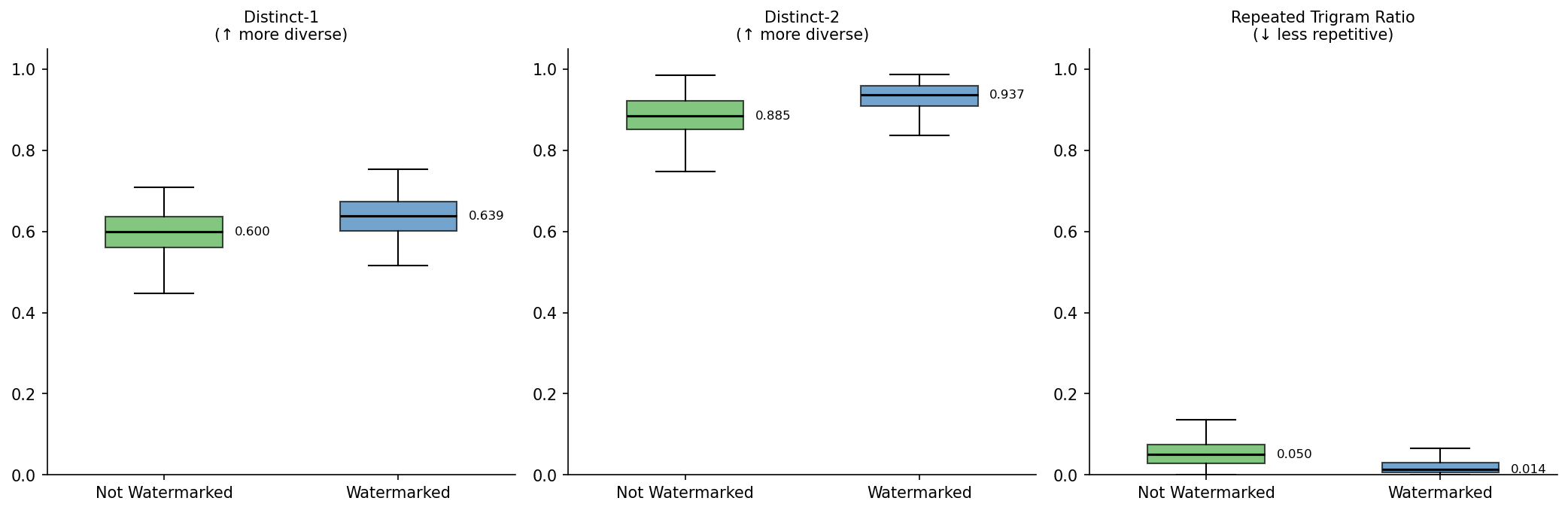}
  \caption{Lexical diversity distributions for 200-token \texttt{Mistral~7B~Instruct} generations on the SAR Prompts dataset~\cite{janas2026sarprompts}.}
  \label{fig:diversity_mistral}
\end{figure*}

We conduct text quality evaluation comparing generated text with and without an embedded watermark. We assess quality along three dimensions: LLM-as-a-Judge pairwise comparison, perplexity, and lexical diversity via word-level distinctness and repetition metrics.

\subsection{LLM-as-a-Judge Pairwise Evaluation}

For this evaluation, texts are generated with \texttt{Mistral~7B~Instruct}. The non-watermarked baseline uses nucleus sampling ($\texttt{temperature}=0.6$, $\texttt{top\_p}=0.9$) with no logit bias, while the watermarked output uses greedy decoding ($\texttt{temperature}=0$) with a logit bias of $\delta = 6$. For each prompt, the judge receives two anonymised responses side-by-side: one watermarked and one non-watermarked. The positions are randomised per-pair to mitigate position bias. The judge model, \texttt{Gemma~4~31B~IT}, is instructed to rate which response is better on the dimension of \textit{social authenticity}---how believable the text is as a genuine human-written product review or social media post. The LLM judge decides a winner (or a tie), preceded by brief reasoning. Texts were generated using SAR Prompts dataset~\cite{janas2026sarprompts}, which contains review-style and social-post prompts.

Table~\ref{tab:llm_judge} reports the results for two codeword lengths ($u=127$ and $u=200$). The two conditions receive very similar ratings, with a slight bias toward the watermarked text at both lengths. The margin is modest---roughly 9~percentage points at $u=127$ and 10~percentage points at $u=200$---and the tie rate is non-negligible, especially at $u=127$.

A plausible explanation for this small difference is that the logit bias pushes the model away from its most probable tokens, introducing slightly more variation in phrasing that the judge occasionally reads as marginally less recognisably AI-generated.

Manual inspection of the judge's reasoning supports this interpretation. In the majority of comparisons the judge describes the two responses as broadly equivalent, often labelling the winner only ``slightly'' better. Across both conditions, the most common criticism is that both texts read like ``generic marketing copy'' or ``AI-generated product descriptions'' rather than genuine human reviews. The watermarked text occasionally wins by adopting a more conversational or first-person tone, while the non-watermarked text is sometimes faulted for overly formal or ``corporate'' phrasing. Overall, the judge perceives the two conditions as close in quality, with the watermark producing a small but consistent edge in perceived authenticity.

\begin{table}[ht]
\caption{LLM-as-a-Judge pairwise comparison on social authenticity. Generation: \texttt{Mistral~7B~Instruct}; judge: \texttt{Gemma~4~31B~IT}. Position randomised per pair.}
\label{tab:llm_judge}
\centering
\setlength{\tabcolsep}{4pt}
\begin{tabular}{@{}lccc@{}}
    \toprule
    \textbf{Length} & \textbf{Not WM} & \textbf{Watermarked} & \textbf{Tie} \\
    \midrule
    $u=127$ & 39.7\% & 49.0\% & 11.3\% \\
    $u=200$ & 43.4\% & 53.7\% & 2.9\% \\
    \bottomrule
\end{tabular}
\end{table}

\subsection{Perplexity Evaluation}

\begin{figure}[ht]
  \centering
  \includegraphics[width=\columnwidth]{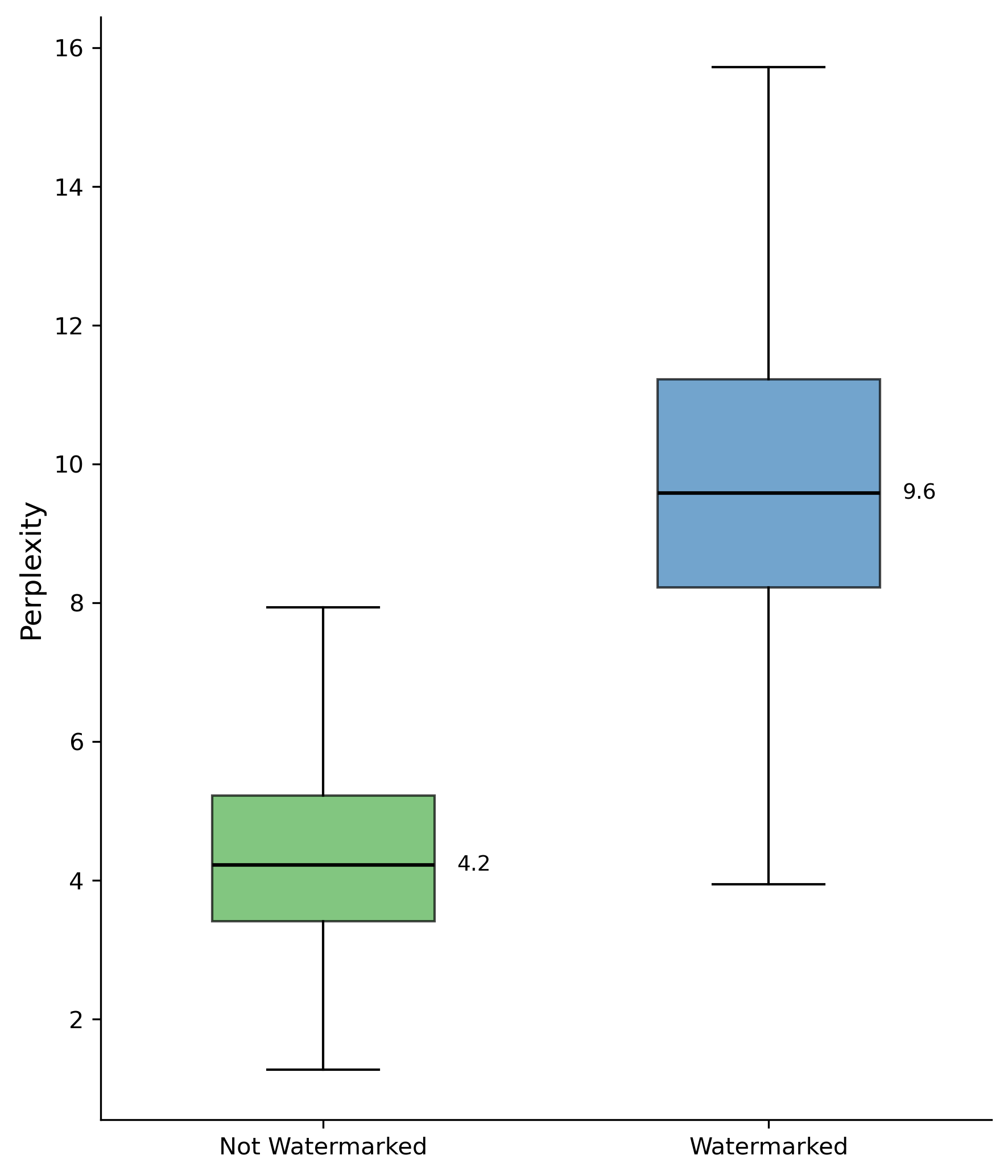}
  \caption{Perplexity distribution for 200-token \texttt{Llama~2~7B} generations on the OpenGen dataset. Evaluator: \texttt{Gemma~4~31B}.}
  \label{fig:perplexity}
\end{figure}

We report perplexity only because it is customary in watermarking papers, but we complement it with other quality metrics, which are more reliable and better suited.
Prior work has shown that likelihood-oriented decoding can produce bland or repetitive text, and that decoding strategy alone can strongly affect text quality. More directly, \cite{wang2023perplexityplmunreliableevaluating} argue that perplexity from pretrained language models is unreliable for evaluating text quality, especially for short texts.

For this evaluation, texts are generated with \texttt{Llama~2~7B}. We evaluate perplexity on 200-token texts on the OpenGen dataset~\cite{opengendataset}. Both conditions produced approximately 7{,}000 generations. Token-level perplexity is computed with a sliding-window approach using the \texttt{Gemma~4~31B} evaluator model. Figure~\ref{fig:perplexity} shows the distributions. Non-watermarked text achieves a median perplexity of 4.22 (mean 4.48), while watermarked text reaches 9.58 (mean 9.91)---roughly a two-fold increase. This increase is expected: the logit bias deliberately pushes the generator away from its most probable tokens, which raises perplexity. The absolute values remain moderate, and the relative increase is comparable to that reported by Qu \emph{et al.}~\cite{qu2025} for the same bias.

\subsection{Diversity Evaluation}

We measure diversity with word-level lexical diversity metrics: Distinct-1 (unique unigrams over total unigrams), Distinct-2 (unique bigrams over total bigrams), and Trigram Repetition Rate ($1 - \text{unique}/\text{total}$ trigrams). Higher Distinct scores indicate more varied vocabulary; lower Trigram Repetition Rate indicates less repetition. All metrics are computed with a lightweight CPU script that tokenises text at word boundaries and requires no model or GPU.

Figures~\ref{fig:diversity_llama2} and~\ref{fig:diversity_mistral} show the distributions for \texttt{Llama~2~7B} ($\approx$7,000 texts per variant) and \texttt{Mistral~7B~Instruct} ($\approx$150 texts per variant), respectively. In both cases the watermarked text is consistently more diverse and less repetitive.

These results confirm that the logit bias does not merely raise perplexity---it actively increases lexical variety. By steering the model away from its most probable tokens, the watermark produces less repetitive, more idiosyncratic phrasing. This explains both the higher perplexity reported above and the slight authenticity edge observed in the LLM-as-a-Judge evaluation.

Randomly selected examples of watermarked generations, together with the prompts that produced them, are provided in Appendix~\ref{sec:generated_examples}.

\section{Generated Text Examples}
\label{sec:generated_examples}

Below are randomly selected examples of watermarked generations, together with the prompts that produced them. Generated text length is 200 tokens.

\subsection{SAR Prompts}
\label{subsec:custom_examples}
Texts generated with \texttt{Mistral~7B~Instruct} on the \textsc{SAR Prompts} dataset~\cite{janas2026sarprompts} (review-style and social-post prompts).

{\small\begin{quote}
\textbf{Prompt:}\\
\textit{Compose a review of the Shimano GRX gravel bike groupset for mixed-terrain riding.}

\textbf{Generation:}\\
The Shimano GRX gravel bike groupset has been designed for those who enjoy mixed-terrain cycling. The GRX groupset features Shimano's latest technology, including their new hydraulic disk brakes, which offer excellent power and accuracy. The groupset also includes Shimano's new 48-11 cassette and cranks, which provide the perfect range of gear ratios for tackling a variety of terrains. The GRX groupset also includes Shimano's Di2 shifting, which offers smooth shifting, even on the most technical terrains.

One of the standout features of the Shimano GRX groupset is its versatility. It is suitable for cycling both gravel paths and pavement roads, making it a great choice for those looking to explore a variety of terrain. Additionally, it is also suitable for cycling in a wide range of weather conditions
\noindent\rule{\linewidth}{0.4pt}
\end{quote}}

{\small\begin{quote}
\textbf{Prompt:}\\
\textit{Write a review of the Anker 737 Power Bank evaluating its 140W output for charging laptops on the go.}

\textbf{Generation:}\\
The Anker 737 Power Bank is a portable charger that is great for charging laptops on the go. With its 140W output power, you are sure that your laptop will be charged to your liking. The power bank is small and convenient to carry, making it perfect for travelers, students and anyone who wants to keep their laptop charged on-the-go.
One of the great things about this power bank is its compatibility with a wide variety of laptops. It is designed to work with laptops that are compatible with USB-C chargers, which includes most of modern laptops today. Additionally, it also has two other ports that can be used to charge two devices at the same time, making it perfect for charging other devices such as your smartphone.
Another thing I like about this power bank is its durability. The power bank is equipped with an advanced battery technology and is rated for 1,500 charge
\noindent\rule{\linewidth}{0.4pt}
\end{quote}}

{\small\begin{quote}
\textbf{Prompt:}\\
\textit{Write a review of the KitchenAid Artisan stand mixer for baking bread and making pasta attachments.}

\textbf{Generation:}\\
The KitchenAid Artisan stand mixer is a high-quality and reliable option for both baking bread and making pasta. The mixer's sturdy and powerful design makes it easy to handle a variety of tasks, from kneading dough to whipping cream. The bowl-lift design also makes it easy to access the bowl and ingredients without the mixer being on, making it a safer option for those who prefer to work with the mixer unplugged.

One great aspect of using the Artisan for baking bread is that you can use any recipe that calls for a bread machine. The Artisan's versatility means you can make everything from simple white bread to complex sourdough bread. The kneading capabilities of the mixer are also impressive. It's easy to switch between dough and bread modes and adjust the time and speed to suit your needs.

In addition to baking bread, the KitchenAid also has many

\noindent\rule{\linewidth}{0.4pt}
\end{quote}}

{\small\begin{quote}
\textbf{Prompt:}\\
\textit{Draft a review of the Bang and Olufsen Beoplay HX headphones for luxury design and soundstage.}

\textbf{Generation:}\\
The Bang and Olufsen Beoplay HX are an exquisite pair of headphones that combine luxury and soundstage in an unrivaled way. The headphone's sleek and elegant design exude a feeling of opulence and sophistication. From the premium leather earpads and headband to the meticulously engineered aluminum frame, the build quality and attention to detail in the Beoplay HX is truly remarkable. The headphones come in several eye-catching colors, including black, gray, and gold. The gold version, in particular, is a true statement piece, with a stunning gold finish and a luxurious texture to the touch.

One of the standout features of the Beoplay HX is its expansive and immersive soundstage. The headphones are powered by a high-performance Bang \& Olufsen proprietary driver, which delivers crystal-clear and detailed sound with a wide
\noindent\rule{\linewidth}{0.4pt}
\end{quote}}

{\small\begin{quote}
\textbf{Prompt:}\\
\textit{Create a review of the Apple AirPods Pro 2 for commuting on noisy subway systems.}

\textbf{Generation:}\\
The AirPods Pro 2 are a top-of-the-line option when it comes to noise-cancelling earphones for commuters. The ActiveNoise Cancellation is particularly useful for those looking to block out unwanted sounds on noisy subway systems.

The earbuds are comfortable and secure, making them a good choice for use during long subway rides or commutes. They stay in your ears well, even with all the noise and motion of a busy subway system.

One great feature of these headphones is that they offer a transparency mode that allows you to hear ambient sounds while still benefiting from some degree of active noise cancellation. This is particularly helpful when it is important to be aware of your surroundings, such as when crossing the street or boarding a train.

Another plus is that you can use the AirPods to take calls and listen to music, all while still enjoying the active noise

\rule{\linewidth}{0.4pt}
\end{quote}}

\subsection{OpenGen Dataset}
\label{subsec:opengen_examples}
Texts generated with \texttt{Llama~2~7B} on the OpenGen dataset~\cite{opengendataset}.

{\small\begin{quote}
\textbf{Prompt:}\\
\textit{British army forces in North America were primarily tied up with the Siege of Boston in 1775. Seeking bases of operations where they had more control, the British planned an expedition to the southern colonies. Major General Henry Clinton, then in Boston, was to travel to Cape Fear, North Carolina, where he would join with largely Scottish Loyalists raised in the North Carolina backcountry, and a force of 2,000 men from Ireland under the command of Major General Charles Cornwallis. The plan was beset by difficulties from the start. The Irish expedition, originally supposed to depart at the beginning of December 1775, was delayed by logistical difficulties, and its 2,500 troops did not depart until February 13, 1776, escorted by 11 warships under the command of Admiral Sir Peter Parker. Clinton left Boston on January 20 with two companies of light infantry, and first stopped at New York City to confer with William Tryon, New York's royal governor. Major General Charles Lee, sent by Major General George Washington to see to the defense of New York, coincidentally arrived there the same day as Clinton. New York was at that time extremely tense; Patriot forces were beginning to disarm and evict Loyalists, and the British fleet anchored there was having difficulty acquiring provisions. Despite this, Clinton made no secret that his final target was in the south. Lee observed that this was "certainly a droll way of proceeding; to communicate his full plan to the enemy is too novel to be credited." This was not even the first notice of the expedition to the colonists; a letter intercepted in December had already provided intelligence that the British were planning to go to the South. Clinton arrived at Cape Fear on March 12, expecting to find the European convoy already there. He met with the royal governors of North and South Carolina, Josiah Martin and William Campbell, and learned that the recruited Scottish Loyalists had been defeated at Moore's Creek Bridge two weeks earlier. Clinton also received pleas for assistance from the royal governor of Georgia, James Wright, who had been arrested, and then escaped to a navy ship. Parker's fleet had an extremely difficult crossing. Battered by storms and high seas, the first ships of the fleet did not arrive at Cape Fear until April 18, and Cornwallis did not arrive until May 3.}

\textbf{Generation:}\\
Clinton, meanwhile, was forced to wait until he received word that Cornwallis was safely in port before departing Cape Fear on April 20, in an unsuccessful effort to capture the Patriot garrison in Charlestown (now called Charles Town) in Charleston, S. C.
Clinton's expedition to South Carolina was a complete success for British forces, but was a disaster in the long run. The Patriotic forces were unable to defend Charlestown, the British took over the town without a fight and the Patriotic government fled to the safety of the countryside. The Patriotic forces were able, in the end, to recapture the city in the Battle of Great Swamp, but the loss of control over South Carolina was a major setback for the colonist cause. In the short term, the British gained control over a significant portion of South and North Carolina, and they were able,
\noindent\rule{\linewidth}{0.4pt}
\end{quote}}

{\small\begin{quote}
\textbf{Prompt:}\\
\textit{Attending school became difficult for Radcliffe after the release of the first Harry Potter film. Some fellow pupils became hostile, though he says they were just trying to "have a crack at the kid that plays Harry Potter" rather than acting out of jealousy. As his acting career began to consume his schedule, Radcliffe continued his education through on-set tutors. He admitted he was not a very good student, considering school useless and finding the work "really difficult". He achieved A grades in the three AS-level exams that he took in 2006, but decided to take a break from education and did not go to college or university. Part of his reasoning was that he already knew he wanted to act and write, and that it would be difficult to have a normal college experience. "The paparazzi, they'd love it", he told Details magazine in 2007. "If there were any parties going on, they'd be tipped off as to where they were." Radcliffe first expressed a desire to act at age five, and, aged ten, he made his acting debut in BBC One's televised two-part adaptation of the Charles Dickens novel David Copperfield (1999), portraying the title character as a young boy. Radcliffe made his film debut in The Tailor of Panama (2001), an American film based on John le Carr\'e's 1996 spy novel, and a moderate commercial success. In 2000, producer David Heyman asked Radcliffe to audition for the role of Harry Potter for the film adaptation of Harry Potter and the Philosopher's Stone, the best-selling book by British author J. K. Rowling. Rowling had been searching for an unknown British actor to personify the character, and the film's director Chris Columbus recalled thinking, "This is what I want. This is Harry Potter" after he saw a video of the young actor in David Copperfield. Eight months later, following several auditions, Radcliffe was selected to play the part.}

\textbf{Generation:}\\
The Harry Potter film franchise has been one of Hollywood's most successful series of all time. It was the first time in history that three consecutive movies had earned more than US \$300 million in box office revenue. In addition to the Harry Potter series of films, Radcliffe starred as the title character of the 2006 film version of the BBC television serial adaptation, The Woman in White. He has starred as the young witch-hunter Jonah in The Sorcerer's Apprentice with actor and comedic actor Jonah Hill and Nicolas Cage. In December 2006, Radcliffe made his Broadway debut as the titular character of Peter Shaffer's one-man show Equus, for which he received positive notices from The New York Times critic, Charles Isherwood who called him a "fearless young actor". In 2006 he won
\noindent\rule{\linewidth}{0.4pt}
\end{quote}}

{\small\begin{quote}
\textbf{Prompt:}\\
\textit{On the first night Lead Belly and his wife Martha spent in the city, they encountered racially discriminatory Jim Crow laws similar to those found in their native Louisiana: most hotels refused to rent rooms to African Americans and the few that would were either full or refused to serve him because he was with a white man (Lomax). Lomax, in some versions of the story described as an unnamed "white friend", offered to let the couple stay for the night in his apartment near the Supreme Court Building. The next morning, Lead Belly awoke to Lomax arguing with his landlord about the presence of a black man, with the landlord threatening to call the police. While in Washington, Lead Belly encountered several other incidents of segregation that are believed to have contributed to the impetus of the song. For instance, when Lead Belly, Lomax, and their wives wanted to go out to dinner together, they discovered that it was impossible for the mixed race group to find a restaurant that would serve them. Lead Belly was told that if he returned later without Lomax, he would be served. In response to one of these incidents, a friend of Lead Belly's, variously identified as either Lomax or Mary Elizabeth Barnicle joked that Washington was a "bourgeois town." Though Lead Belly did not know what the word "bourgeois" meant, he was fascinated by the sound of it, and after its meaning was explained to him he decided to incorporate it into a song about the trip. The song came together quickly; one account claims that it only took a few hours for Lead Belly to write it. Lomax liked it because it was partly based on what happened in his apartment. "The Bourgeois Blues" is a blues-style protest song that criticizes the culture of Washington, DC. It protests against both the city's Jim Crow laws and the racism of its white population. Its structure includes several verses and a refrain that declares that the speaker is going to "spread the news all around" about the racial issues plaguing the city. The song, particularly in the refrain, conflates race and economics by referring to Caucasians as "bourgeois".}

\textbf{Generation:}\\
The lyrical structure of the refrain is based upon that of "The House of the Rising Sun" by English musician Eric B. King (1898–?). Lead Belly's version is the only known instance of the song being performed with a refrain, and it has since been performed with the lyrics of "Birmingham Sunday", a song written by Hudie Ledbetter (Lead Belly) in the 1860s. In addition to its lyrical structure, "The Bourgeois Blues" is also musically similar to "The Rising of the Sun" (or the "Rising Sun" Blues) written in 1899. Both songs have a twelve bar form, a similar melancholy feel in the verses that contrast with the more optimistic refrains, a descending chromatic bass line, a descending chromatic bass pattern, and a similar instrumentation consisting only of
\noindent\rule{\linewidth}{0.4pt}
\end{quote}}

{\small\begin{quote}
\textbf{Prompt:}\\
\textit{After two straight second-place finishes at Charlotte and Riverside, Richmond started the Pocono event in third place inside the second row. That race saw a caution for rain with five laps left before the halfway point. NASCAR wanted the cars to get to the halfway point to make the race official, so the sanctioning body had the drivers slowly circle the track. It took the drivers 26 minutes to complete the laps, and the rain was so heavy that some drivers had to look out their side windows because they could not see out their windshields. Two hours later, the track had dried and the race resumed with Richmond in third. After Richmond's car was adjusted to remove the "push", the car was more to his liking. Because his radio did not work, he was unable to communicate with his crew chief, Hyde, and he made his final pit stop with 37 laps left. Hyde worried that Richmond had stopped a lap too early to ensure that he would have enough fuel to make it to the end. After Richmond took the lead with 30 laps left in the race, Dale Earnhardt made up three seconds on Richmond's five-second lead. With four laps to go, Buddy Arrington spun in a three-car accident. The remaining laps of the race where completed slowly under caution and Richmond took the checkered flag for the victory. He had led 97 laps, including the final 30, taking his first victory in a Rick Hendrick car. The tour returned to Pocono a month later, and Richmond battled for another victory in a fog-shortened event. In the final 8-lap sprint, Richmond competed in a three-car battle with Geoff Bodine and Ricky Rudd. Richmond crossed the finish line beside Rudd, winning the race by 0.05 seconds. He notched four more victories that season, and over a span of twelve races, Richmond earned three second-place finishes, and six wins. The National Motorsports Press Association named him Co-Driver of the Year with Earnhardt after Richmond accumulated 13 top 5 finishes and 16 in the top 10. He had a career-best third-place finish in points after winning seven events in 1986, in what was his last full NASCAR season.}

\textbf{Generation:}\\
He was the first driver to win the Winston Millions, a 34-race series, in the 30-race schedule.

\#\# 1993 and later years

Richmond was named the Grand Marshal for the inaugural 1993 Daytona 500. In his final race as the driver for RCR, Richmond started the race in 12th, but had to pit with a broken axel on lap 35. His car returned on Lap 54, and he had to make up 30 positions to get to a position in which he was able to race for positions at the front. He had a strong run, and finished in the top 5, but was disallowed a top five finish after the crew chief made an unapproved pitstop. He was released from RCR after that race, and he was hired as a commentator for TNN. In
\noindent\rule{\linewidth}{0.4pt}
\end{quote}}

{\small\begin{quote}
\textbf{Prompt:}\\
\textit{As a result, it was slowed down. The gameplay originated with a tech demo by Naka, who developed an algorithm allowing a sprite to move smoothly on a curve by determining its position with a dot matrix. Naka's prototype was a platform game with a fast-moving character rolling in a ball through a long, winding tube, and this concept was fleshed out with Ohshima's character designs and levels by Yasuhara. Yasuhara originally intended to work on the game for three months due to the delay of his planned move to the United States by the outbreak of the Gulf War, but was engrossed in the project for nearly a year. His designs for levels were intended to attract both hardcore and casual gamers by integrating occasional challenging set pieces into the mostly accessible level design. The color scheme was influenced by the work of pop artist Eizin Suzuki, and the aesthetics of Green Hill were influenced by the geography of California. In designing the gameplay, Naka was inspired by Mario creator Shigeru Miyamoto, whose games he had enjoyed playing years earlier. Admiring the simplicity of Miyamoto's mechanics in complex environments, Naka decided that Sonic would be controlled with only a directional pad for movement and a single button for jumping. He also wanted his creation to be more action-oriented than the Mario series; while playing Super Mario Bros., he had wondered why the levels could not be cleared more quickly. Naka, Ohshima, and Yasuhara worked 19 hours a day on the project for several months. Due to the need to demonstrate the Genesis' technological prowess, the game underwent extensive testing and redesign, which took over six months. According to Naka, the game had the fastest-ever character speed in a video game and a rotation effect in the special stages that had been considered impossible on the console. The team intended to add a two-player mode displayed via split-screen, but Naka's programming knowledge was insufficient to implement it. A two-player mode appeared in Sonic the Hedgehog 2 (1992), whereby the second player controls Sonic's sidekick Miles "Tails" Prower.}

\textbf{Generation:}\\
The soundtrack, which was composed and produced over a period of three weeks, was composed by Masato Nakamura, who was given a month to complete it after the game went gold. The music was inspired in part by rock musician Billy Corgan and the band he founded in the United States. Nakamura's work was recorded in two weeks with a 200-man orchestra, a 56-man choir, and several solo musicians. Nakamura's compositions were influenced in style by the music of video game soundtracks he had heard in the past; his favorite was Super Castlevania IV's (1989). The sound effects were created in two weeks with the use of the SGI Indigo workstation, a computer with 16 megabytes of RAM, 160 megabytes of disk space, a CD-ROM, and a hard drive, which was unhe
\noindent\rule{\linewidth}{0.4pt}
\end{quote}}

\section*{Acknowledgment}
In accordance with IEEE policy, the authors disclose the use of AI assistants from the GPT (OpenAI) and Claude (Anthropic) model families for software implementation and editing prose. All content was reviewed and verified by the authors.

\bibliographystyle{IEEEtran}
\bibliography{references}

\end{document}